\documentclass{aa}  

\usepackage{graphicx}
\usepackage{txfonts}
\usepackage{amssymb}
\usepackage{multirow}
\usepackage{float}
\usepackage{gensymb}
\usepackage{xcolor}
\usepackage{verbatim} 
\usepackage{rotating}
\usepackage[]{hyperref}
\hypersetup{backref=true, pagebackref=true, hyperindex=true, breaklinks=true,colorlinks=true,urlcolor=blue, linkcolor=blue, citecolor=blue,pagecolor=red, bookmarks=true, bookmarksopen=true}
\begin{document}

   \title{The centimeter emission from planet-forming disks in Taurus}


   \author{A. Garufi\inst{\ref{IRA}, \ref{MPIA}}
          \and C. Carrasco-Gonz\'alez \inst{\ref{Morelia}}
          \and E. Mac\'ias \inst{\ref{ESO_Garching}}
          \and L. Testi \inst{\ref{UniBo}}
          \and P. Curone \inst{\ref{UniMi}, \ref{U_Chile}}
          \and L. Ricci \inst{\ref{CSUN}}
          \and S. Facchini \inst{\ref{UniMi}}
          \and \\ F. Long \inst{\ref{Tucson}, \ref{feng}}
          \and C.F. Manara \inst{\ref{ESO_Garching}}
          \and I. Pascucci \inst{\ref{Tucson}}
          \and G. Rosotti \inst{\ref{UniMi}}
          \and F. Zagaria \inst{\ref{Cambridge}}
          \and C. Clarke \inst{\ref{Cambridge}}
          \and G.J. Herczeg \inst{\ref{Kavli},\ref{PKU}}
          \and \\ A. Isella  \inst{\ref{Rice}}
          \and A. Rota \inst{\ref{Leiden}}
          \and K. Mauc\'o \inst{\ref{ESO_Garching}}
          \and N. van der Marel \inst{\ref{Leiden}}
          \and M. Tazzari \inst{\ref{Arcetri}}
          }

    \institute{INAF - Istituto di Radioastronomia, Via Gobetti 101, I-40129, Bologna, Italy \label{IRA}
   \\
              \email{antonio.garufi@inaf.it}
   \and
   Max-Planck-Institut f\"{u}r Astronomie, K\"{o}nigstuhl 17, 69117 Heidelberg, Germany \label{MPIA}
    \and
    Instituto de Radioastronom\'{i}a y Astrof\'{i}sica (IRyA-UNAM), Morelia, Michoac\'{a}n 58089, Mexico \label{Morelia}
    \and
    European Southern Observatory, Karl-Schwarzschild-Strasse 2, D-85748 Garching, Germany \label{ESO_Garching}
    \and
    Dipartimento di Fisica e Astronomia Augusto Righi, Universit\`a di Bologna, Viale Berti Pichat 6/2, Bologna, Italy \label{UniBo}
    \and
    Dipartimento di Fisica, Universit\`a degli Studi di Milano, Via Celoria 16, 20133 Milano, Italy \label{UniMi}
    \and
    Departamento de Astronom\'{i}a, Universidad de Chile, Camino El Observatorio 1515, Las Condes, Santiago, Chile \label{U_Chile}
    \and
    Department of Physics and Astronomy, California State University Northridge, 18111 Nordhoff Street, Northridge, CA 91330, USA \label{CSUN}
    \and
    Lunar and Planetary Laboratory, University of Arizona, Tucson, AZ 85721, USA \label{Tucson}
    \and
    NASA Hubble Fellowship Program Sagan Fellow \label{feng}
    \and
    Institute of Astronomy, University of Cambridge, Madingley Road, Cambridge CB3 0HA, UK \label{Cambridge}
    \and
    Kavli Institute for Astronomy and Astrophysics, Peking University, Yiheyuan 5, Haidian Qu, 100871 Beijing, China \label{Kavli}
    \and
    Department of Astronomy, Peking University, Yiheyuan 5, Haidian Qu, 100871 Beijing, China \label{PKU}
    \and
    Department of Physics and Astronomy, Rice University, 6100 Main Street, MS-108, Houston, TX 77005, USA \label{Rice}
    \and
    Leiden Observatory, Leiden University, P.O. Box 9513, 2300 RA Leiden, The Netherlands \label{Leiden}
    \and
    INAF, Osservatorio Astrofisico di Arcetri, Largo Enrico Fermi 5, I- 50125 Firenze, Italy \label{Arcetri}
    }

   \date{Received -; accepted -}

 
\abstract
{The last decade has witnessed remarkable advances in the characterization of the (sub-)millimeter emission from planet-forming disks. Instead, the study of the (sub-)centimeter emission has made more limited progress, to the point that only a few exceptional disk-bearing objects have been characterized in the centimeter regime. This work takes a broad view of the centimeter emission from a large sample with VLA observations that is selected from previous ALMA surveys of more representative disks {in brightness and extent}. We report on the detection and characterization of flux at {centimeter wavelengths from} 21 sources in the Taurus star-forming region. Complemented by literature and archival data, the entire photometry from 0.85 mm to 6 cm is fitted by a two-component model that determines the ubiquitous presence of free-free emission entangled with the dust emission. The {flux density of the} free-free emission is found to scale with the accretion rate but is independent of the outer disk morphology depicted by ALMA. The dust emission at 2 cm is still appreciable, and offers the possibility to extract an unprecedented large set of dust spectral indices in the centimeter regime. A pronounced change between the median millimeter indices (2.3) and centimeter indices (2.8) suggests that a large portion of the disk emission is optically thick up to 3 mm. The comparison of both indices and fluxes with the ALMA disk extent indicates that this portion can be as large as 40 au, and suggests that the grain population within this disk region that emits the observed centimeter emission is similar in disks with different size and morphology. All these results await confirmation and dedicated dust modeling once facilities like ngVLA or SKA-mid are able to resolve the centimeter emission from planet-forming disks and disentangle the various components.}
  

   \maketitle
%

\section{Introduction}
Much progress has been made in the characterization of planet-forming disks over the last decade. The advent of the Atacama Large (sub-)Millimeter Array (ALMA) \citep[see e.g.,][]{ALMA2015} and of the extreme adaptive optics facilities \citep[see review by][]{Benisty2023} has enabled the imaging of the circumstellar disks around Class II objects \citep{Lada1987} with high resolution (less than 0.1\arcsec) and high contrast (of the order of 10$^{-5}$). One of the benefits of this type of observations is the possibility to detect and resolve the bulk population of planet-forming disks, namely those disks with an extent of few tens of au (0.2\arcsec -- 0.3\arcsec\ at the distance of the closest star-forming regions). The reason why the community aimed at the characterization of the inner few tens of au is twofold. On the one hand, planet formation via core accretion is favored in this disk region \citep[e.g.][]{Pollack1996}. On the other hand, a less biased view of planet-forming disks (switching from the characterization of extremely large disks to that of ordinary objects) is necessary to bridge the fields of the planet formation and of exoplanets \citep[see e.g.,][]{Mulders2021, vanderMarel2021b, Miotello2023}.

The main finding on the bright and extended planet-forming disks that have been surveyed by sub-millimeter and near-IR high-resolution images is the high occurrence of disk sub-structures \citep[see e.g.,][]{Andrews2018, Garufi2018}. Different explanations have been put forward to explain the origin of these sub-structures \citep[see review by][]{Bae2023}, such as the interaction with planets \citep[see e.g.,][]{Kley1999, Crida2006} or the presence of instabilities \citep[see review by][]{Lesur2023}. The first type of sub-structures to be studied were the disk cavities \citep[pioneerly imaged by e.g.,][]{Dutrey1994, Fukagawa2006, Brown2009, Andrews2011}. Cavities are revealed as a large inner portion (few to several tens of au) of the disk with a diminished dust emission. Cavity-hosting disks (often referred to as transition disks) are however a very peculiar path (or stage) of the disk evolution, and this type of disks only represents 10\% or less of the total disk population \citep[see review by][and references therein]{vanderMarel2023}. In view of this, much effort has been made to improve our understanding of full disks (here meant in opposition to transition disks) and the morphology of other possible sub-structures such as rings and gaps. 

When observed in near-IR scattered light, full disks appear much fainter than cavity-hosting disks because the inner portion of the disk at (sub-)au scale casts a shadow on the outer, resolvable disk \citep{Dullemond2004a, Garufi2017, Garufi2022b}. When observed by ALMA, the inner region of full disks tends to appear bright and smooth \citep[see e.g.,][]{Andrews2018, Long2019}. The correlation between the disk size and luminosity from millimeter images suggests that the emission from a significant portion of the disk may be optically thick \citep{Tripathi2017, Tazzari2021a}. Also, dust scattering can lower the disk emission thus as to decrease the measured optical depth of possibly optically thick emission \citep{Birnstiel2018, Zhu2019b, Sierra2020}. Optically thick emission can also be constrained from the occultation of background material \citep[such as a back-side outflow cavity, see][]{Garufi2020a} or from the CO brightness decrease in proximity of continuum rings \citep{Isella2018, Guzman2018}. The possibility that some disk emission is optically thick poses severe limitations toward the study of the dust grain properties since a high optical depth can conceal the actual dust density and alter the measured spectral indices in a similar manner to the growth of particles to millimeter sizes. 

The (sub-)centimeter emission from disks (frequencies $\nu$ $\sim$ 10--100 GHz) is therefore an essential tool to probe dust densities and grain properties because the optical depth is sufficiently low while the disk emission is still strong enough to be detected in a reasonable amount of telescope time. While the current generation of radio interferometers does not offer the resolving power of ALMA, these may still provide pivotal constraints on the dust properties through unresolved images, or moderately resolved images at the higher frequencies for large and bright disks. The Karl G. Jansky Very Large Array (VLA) and the Australia Telescope Compact Array (ATCA) have produced promising results in this direction focusing on bright disks such as those of LkCa15 \citep{Isella2014}, GM Aur \citep{Macias2016, Macias2018}, MWC758 \citep{Marino2015b, Casassus2019}, HD169142 \citep{Macias2017}, HL Tau \citep{Carrasco-Gonzalez2016, Carrasco-Gonzalez2019}, {HD142527 \citep{Casassus2015}, DM Tau \citep{Liu2024},} or HD163296 \citep{Guidi2022}, as well as on 15 disks with known large cavities \citep{Norfolk2021}. 

In this study, we move the focus from the (sub-)centimeter emission of extraordinary, individual objects to that of a large sample of ordinary disks imaged with VLA in the Taurus star-forming region. The Taurus-Auriga complex is one of the closest star-forming regions \citep{Galli2018} hosting a few hundred low-mass stars \citep{Esplin2019} at diverse evolutionary stages \citep{Kenyon2008}. With more than 60 members with high-resolution images of the disk available, it is among the most studied star-forming regions. The ALMA and near-IR surveys of Taurus from the literature are extensively described in Sect.\,\ref{sec:sample} along with the new VLA dataset. Then, in Sect.\,\ref{sec:results} we analyze the VLA observations focusing on the unresolved information. At centimeter wavelengths, non dust emission (e.g., free-free and synchrotron emission) plays an important role. Therefore, the images are modeled to disentangle the photometry from the dust emission and that from non dust emission (Sects.\,\ref{sec:overview} and \ref{sec:sed}). The non dust emission is analyzed in Sect.\,\ref{sec:free-free} while the dust emission in Sects.\,\ref{sec:dust_indices} and \ref{sec:dust_masses}. Finally, we discuss our results and conclude in Sects.\,\ref{sec:discussion} and \ref{sec:conclusions}.

\section{Sample and observing setup} \label{sec:sample}
The sample analyzed in this work consists of 21 sources in Taurus with VLA observations available. In this section, we describe the stellar and disk properties known from the literature (Sect.\,\ref{sec:targets}) and present the new (sub-)centimeter observations from VLA (Sect.\,\ref{sec:observations}) {as well as some complementary observations from CARMA (Sect.\,\ref{sec:observations_CARMA})}.

\subsection{Target properties} \label{sec:targets}
All the sources in this work are part of the ALMA survey described by \citet{Long2018,Long2019}. Their original sample was compiled from the whole known census of Taurus excluding only stars with type later than M3 as well as close binary system and highly extincted sources. In addition, they avoided all disks with existing high-resolution ALMA images at that time. This was the main bias of the selection as it excludes all well-studied (and often extraordinarily bright and extended) disks such as those of AB Aur, GM Aur, and GG Tau.  

Within their original sample, \citet{Long2019} revealed 12 disks with evident dust gaps and rings (hereafter sub-structured disks) and 20 disks with no resolvable sub-structures. The sub-structured disks have an effective dust radius with 95\% flux encircled larger than 50 au while the other disks are always smaller than that (motivating the definition hereafter of compact disks). The inner part of sub-structured disks show similar brightness properties to the compact disks, making the existence of sub-structures beyond $\sim$50 au to be the key difference between the two categories. Our sample comprises 10 sub-structured disks and 11 compact disks, as is shown in Table \ref{table:properties}, where their effective radius is also shown\footnote{Due to the conversion to the intrinsic spatial scale, two sub-structured disks (around IP and FT Tau) are actually smaller than the largest compact disk (in DR Tau).}. 

Of the 21 sources analyzed in this work, 18 have also been observed in near-IR scattered light with VLT/SPHERE \citep{Garufi2024}. All these disks appear relatively faint in scattered light. This reflects the aforementioned absence from the sample by \citet{Long2019} of extraordinarily large disks with large cavities that are typically bright in scattered light \citep[see][]{Garufi2017}. However, 3 of the 18 targets observed with SPHERE (DO, DR, and HP Tau) show evidence of ambient material (envelopes, outflows, streamers). This material is mostly composed of interstellar gas and $\mu$m-sized dust grains that escape the continuum surveys by ALMA.  

The stellar properties of the sample are also listed in Table \ref{table:properties}. Most of the sources are sub-solar-mass stars with the only exceptions being HP Tau, HQ Tau, and MWC 480. The ages are typically in the 1--3 Myr interval as expected for the Taurus star-forming region \citep[see][for details]{Garufi2024}. The mass accretion rates substantially vary within the sample spanning from $2.5 \cdot 10^{-8}$ M$_\odot$ yr$^{-1}$ (DL Tau) to $2.5 \cdot 10^{-11}$ M$_\odot$ yr$^{-1}$ (HP Tau).

\begin{table}
\caption{Main properties of the sample from the literature.}             
\label{table:properties}
\centering              
\begin{tabular}{l c c c c c}  
\hline\hline              
Target & $d$& $M_*$ & log(${\dot M}_{\rm acc}$) & Disk & $R_{\rm ALMA}$  \\ 
 & (pc) & (M$_\odot$) & (M$_\odot$ yr$^{-1}$) & type & (au) \\
\hline                     
   BP Tau & 127.4 & 0.4 & -8.2 & $\cdot$ & 41 \\  
   CIDA 9 & 175.1 & 0.5 & -8.7 & $\odot$ & 65 \\ 
   DL Tau & 159.9 & 0.6 & -7.6 & $\odot$ & 165 \\  
   DN Tau & 128.6 & 0.5 & -9.0 & $\odot$ & 61 \\ 
   DO Tau & 138.5 & 0.5 & -7.8 & $\cdot$ & 37 \\  
   DQ Tau & 195.4 & 0.5+0.5 & -7.7 & $\cdot$ & 43 \\ 
   DR Tau & 193.0 & 0.4 & -7.9 & $\cdot$ & 53 \\  
   DS Tau & 158.4 & 0.5 & -8.6 & $\odot$ & 71 \\ 
   FT Tau & 130.2 & 0.3 & -8.9 & $\odot$ & 46 \\  
   GI Tau & 129.4 & 0.7 & -8.4 & $\cdot$ & 25 \\ 
   GK Tau & 129.1 & 0.9 & -8.7 & $\cdot$ & 13 \\  
   GO Tau & 142.4 & 0.4 & -9.1 & $\odot$ & 169 \\ 
   HO Tau & 164.5 & 0.3 & -9.1 & $\cdot$ & 44 \\  
   HP Tau & 171.2 & 1.7 & -10.6 & $\cdot$ & 21 \\ 
   HQ Tau & 161.4 & 1.3 & -8.4 & $\cdot$ & 25 \\  
   IP Tau & 129.4 & 0.6 & -9.1 & $\odot$ & 36 \\ 
   IQ Tau & 131.5 & 0.5 & -8.5 & $\odot$ & 110 \\  
   MWC 480 & 156.2 & 2.1 & -7.6 & $\odot$ & 137 \\  
   UZ Tau E & 123.1 & 0.4+0.3 & -8.0 & $\odot$ & 82 \\ 
   V409 Tau & 129.7 & 0.4 & -9.6 & $\cdot$ & 40 \\  
   V836 Tau & 167.0 & 0.3 & -8.4 & $\cdot$ & 31 \\ 
\hline 
\end{tabular}
\tablefoot{CIDA 9 can be identified as IRAS 05022+2527 while HP Tau as V$^*$ HP Tau. MWC 480 is also known as HD31648. Columns are target name, distance, stellar mass, mass accretion rate, disk morphology (\mbox{$\cdot$ = compact}, \mbox{$\odot$ = sub-structured}) from 1.3 mm continuum emission \citep{Long2019}, and effective disk radius at 1.3 mm from ALMA. The distance is from \textit{Gaia} DR3 \citep{Gaia2023}, the stellar mass from various sets of pre-main-sequence tracks is from \citet{Garufi2024}, the mass accretion rate from \citet{Gangi2022}, except GO Tau and HO Tau from \citet{Ricci2010}, CIDA 9 from \citet{Harsono2024} and HP Tau from \citet{Lin2023}, while the ALMA radius is from \citet{Long2019}, obtained from their angular measurement of the effective radius encircling 95\% of the flux multiplied by the distance from \textit{Gaia}. The double stellar masses of DQ Tau and UZ Tau E indicates spectroscopic binaries. }
\end{table}

\subsection{VLA observations and data reduction} \label{sec:observations}
We make use of VLA observations of the Taurus star-forming region in the Q-band ({bandwidth: 39--47 GHz}, $\lambda_{\rm c}$=7 mm), Ka-band ({29--37 GHz}, $\lambda_{\rm c}$=1 cm), K-band ({18--26 GHz}, $\lambda_{\rm c}$=1.3 cm), Ku-band (12--18 GHz, $\lambda_{\rm c}$=2 cm), and C-band (4--8 GHz, $\lambda_{\rm c}$=6 cm). All observations are from projects 20A-373 (PI: M. Tazzari) and 21B-267 (PI: C. Carrasco-Gonzalez). All sources are observed in the Ku band and nineteen of them also in the Ka. Instead, only nine sources are observed in the Q and K bands, and ten sources in the C band. The observations in the Q, Ka, and K bands were taken in B array configuration resulting in a typical beam size of 0.15\arcsec, 0.25\arcsec, and 0.35\arcsec, respectively, while those in the Ku and C bands were taken in the C configuration yielding a beam size of approximately 1.4\arcsec\ and 3\arcsec, respectively. The typical RMS of the observations are of the order of a few $\mu$Jy. The details of each individual dataset are given in Table \ref{table:setup}.

Data were first calibrated by the VLA pipeline. Secondly, phase self-calibration was performed on {all clear detections} using the software \textsc{CASA}, version 6.4. Each dataset was spectrally averaged and shifted to a common center across multiple wavebands. The cleaning procedure was performed employing a Briggs weighting of 0.5 and making sure that even background sources at the edge of the field of view (at a scale of several arcminutes) were properly masked. The lack of this operation would in fact result in a dramatic increase for the flux of the central source after the first round of self-calibration (up to 4--5 times the initial value). Instead, the images obtained after one round of self-calibration performed {with the largest possible solution interval and with a minimum SNR of 2} for each polarization independently and combining across scans exhibit a central flux that is typically 10\% higher than the original value with only a few cases up to 20\%. Further rounds of self-calibration {with shorter solution intervals or combining across spectral windows} were always unnecessary to recover any additional flux. {We find no correlation between the amount of flux recovered and the initial brightness of the source.}

\subsection{CARMA observations and data reduction} \label{sec:observations_CARMA}

{We also make use of unpublished CARMA data of 7 sources from the CARMA data archive\footnote{\url{http://carma-server.ncsa.uiuc.edu:8181/}} that will be described in depth by Blair et al.\,in prep. The inclusion of these data is motivated by the poor available photometry from the literature at 3 mm (see Sect.\,\ref{sec:sed}) with only 13 of our 21 sources having published 3mm-photometry.} 

{The CARMA observations in question consists of 100--113 GHz images of the entire sample (except CIDA 9, UZ Tau and V409 Tau) that were taken between October 2009 and December 2012. The images were acquired using the sub-array with 15 antennas in C, D, and E configurations resulting in a minimum angular resolution of 1.8\arcsec. The data were calibrated and optimized with the MIRIAD software as is described in detail by Blair et al.\,in prep.}

\section{Results} \label{sec:results}
In this section, we analyze the new VLA observations. First, an overview of the VLA images is given in Sect.\,\ref{sec:overview}. Then, the spectral energy distributions of the sample are modeled with two power-laws (Sect.\,\ref{sec:sed}) and the results from non dust emission are presented in Sect.\,\ref{sec:free-free}. After that, the dust emission is analyzed in Sects.\,\ref{sec:dust_indices} and \ref{sec:dust_masses}.

\subsection{Overview of the VLA images} \label{sec:overview}
All sources are detected in the Q, Ka, K, and Ku band when the respective data is available. Instead, only two sources (BP Tau and HQ Tau) are detected in the C band, with the latter being a very marginal detection. The signal in all Ku-band images is unresolved while it is marginally resolved in some Ka-band images and clearly resolved in most Q-band images. An investigation of the resolved component of the Q- and Ka-band images is deferred to a shortcoming study. In this work, we focus on the integrated flux at Q-band through C-band putting particular emphasis on the Ku-band data (2 cm) for which very few studies have been carried out in the past. All images in the Ku band are shown in Fig.\,\ref{fig:images_ku}.

The total flux associated with each source and waveband was measured through a gaussian fit performed with the CASA task \textsc{imfit}. 
The measured fluxes are reported in Table \ref{table:setup}. These typically span from a few mJy at the higher frequencies to a few tens of $\mu$Jy at the lower frequencies. Two datasets represent a notable exception: the Ku-band flux of HP Tau (7 mJy that is two orders of magnitude higher than the other sources) and the K-band flux of DQ Tau (31 mJy). These two measurements are discussed in Sect.\,\ref{sec:discussion_cases}.

\begin{figure}
   \includegraphics[width=9cm]{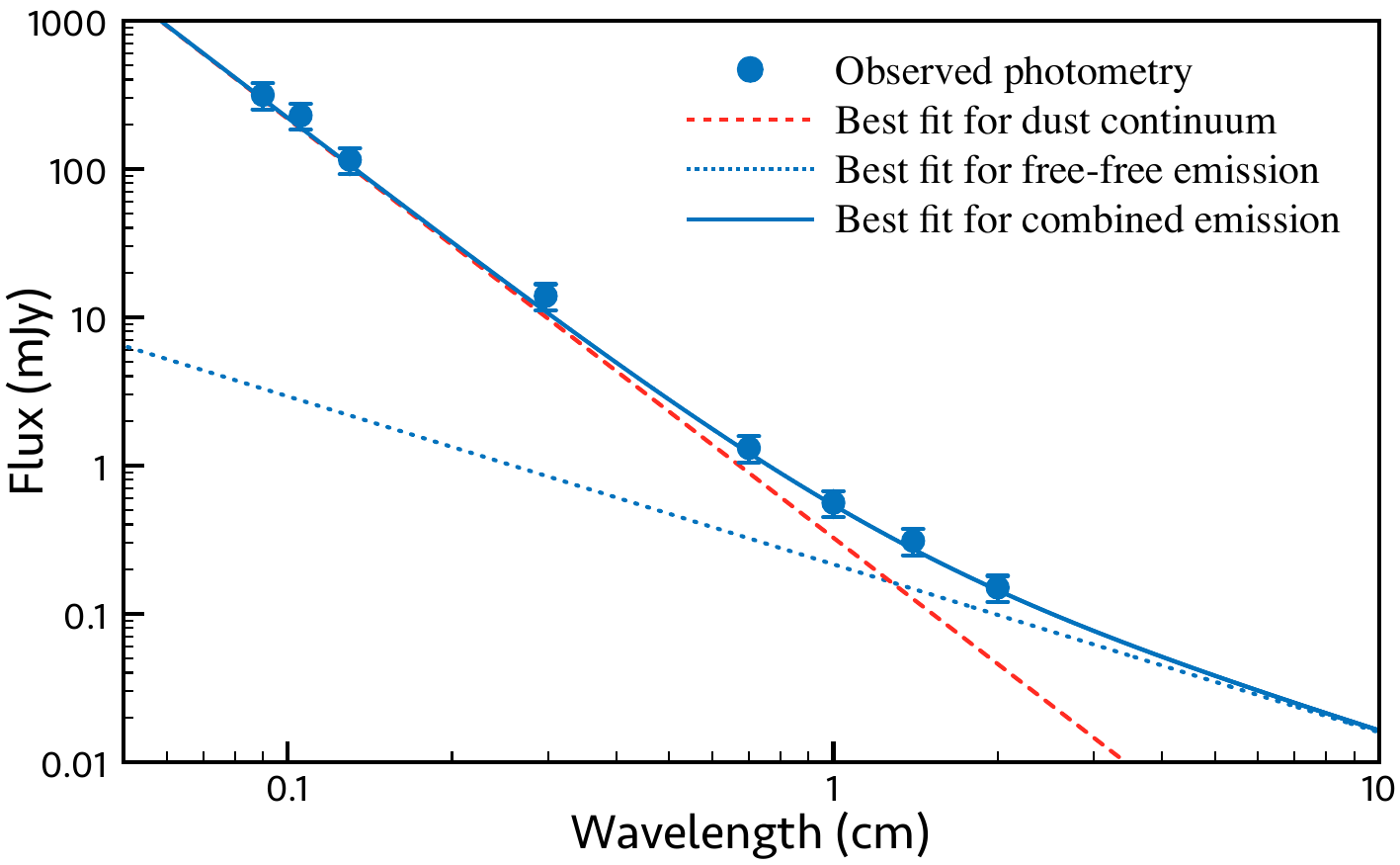}
   \caption{Illustrative SED. The observed photometry of DR Tau is fit by two power laws, for the dust continuum and the free-free emission. All derived slopes and all SEDs are shown in Table \ref{table:sed_fit} and Fig.\,\ref{fig:all_seds}, respectively.}
              \label{fig:sed_illustrative}%
    \end{figure}

\subsection{Spectral energy distribution} \label{sec:sed}
The VLA {and CARMA} photometry of each source was then placed in a spectral energy distribution (SED) together with the photometry at 850 $\mu$m from SCUBA \citep{Andrews2005}, at 1.06 mm from CSO \citep{Beckwith1991}, at 1.33 mm from ALMA \citep{Long2019}, and at 2.7--3.5 mm from the IRAM Plateau de Bure Interferometer \citep{Dutrey1996, Ricci2010}.

A first visual inspection of the SEDs reveal a clear flattening of the trend for the photometry at wavelengths longer than 1 cm. This suggests the presence of an extra radio emission component such as free-free emission from jets \citep[e.g.,][]{Anglada2018} and photo-evaporating wind \citep{Pascucci2012}, or gyro-synchrotron radiation from magnetically induced stellar flares \citep{Dulk1982}. A useful tool to disentangle between free-free and gyro-synchrotron emission is the spectral slope of the associated SED, with the former emission producing positive slopes with the frequency \citep{Panagia1975, Reynolds1986} and the latter producing negative slopes from $\sim$5 GHz on \citep{Bastian1998}, that are positive trends with the wavelength up to $\sim$5 cm. The systematic non-detection of any flux in our C-band images (at 6 cm) and the lower fluxes recorded in the Ku  (2 cm) compared to the Ka (1 cm) intuitively advocate for a free-free emission being responsible for the extra radio emission.      
\begin{figure}
   \includegraphics[width=9cm]{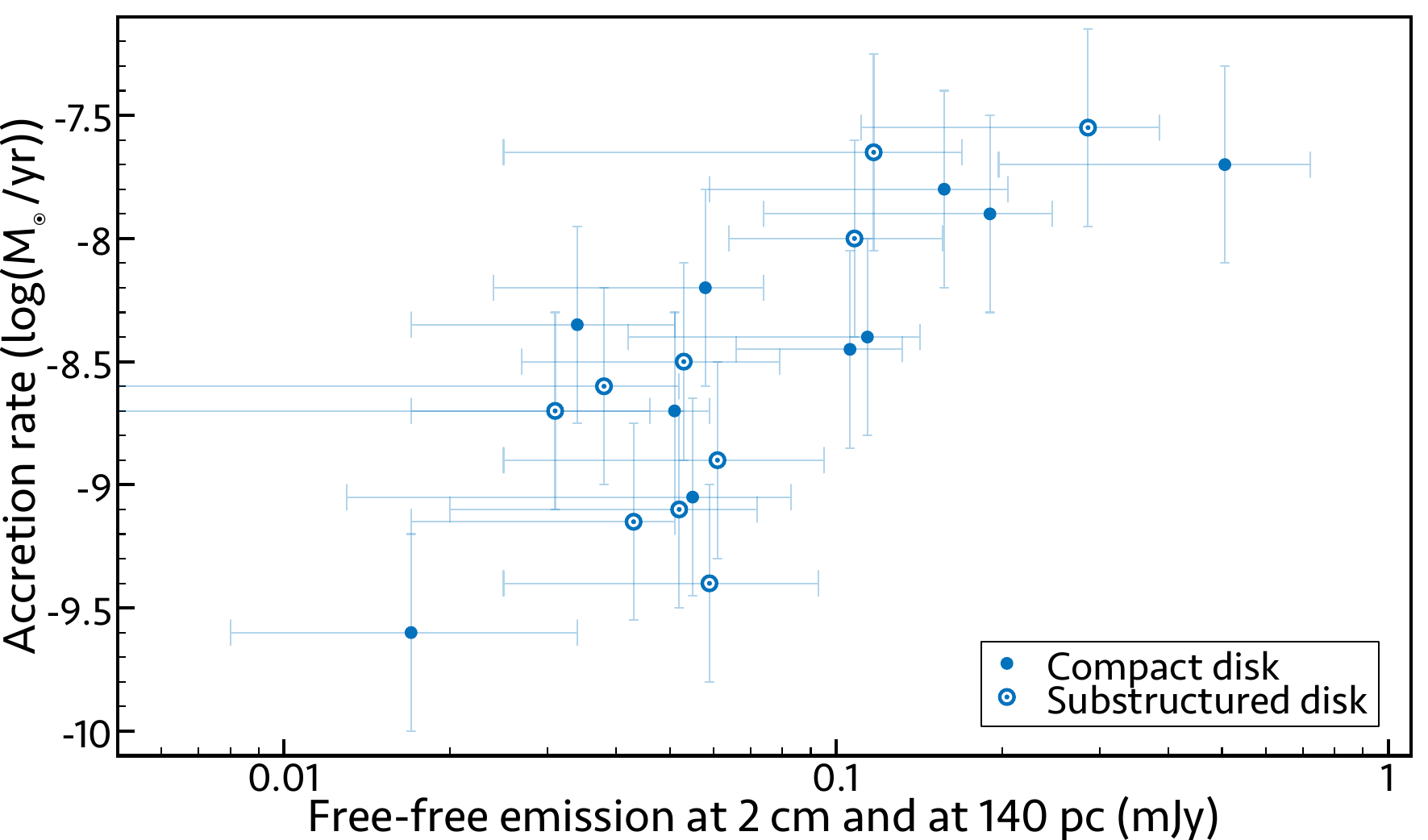}
   \caption{Free-free emission vs mass accretion rate. The free-free emission at 2 cm constrained by our model (see Sect.\,\ref{sec:sed}) for the entire sample is scaled to 140 pc and compared with the mass accretion rates shown in Table \ref{table:properties}. Despite large errors, a correlation between both quantities is clearly seen across several orders of magnitude. Nonetheless, the distribution of compact disks against substructured disks appears random. HP Tau (with a free-free emission at 140 pc of approximately 10 mJy and an accretion rate of $2.5 \cdot 10^{-11}$ M$_\odot$ yr$^{-1}$) clearly falls out of the correlation, and is not shown for a better view of the plot.}
              \label{fig:acc-ff}%
    \end{figure}

To quantify the contribution from the {ionized gas} in our sources (free-free emission {or gyro-synchrothron emission}), each SED was therefore fitted with a simple model consisting on the sum of two power-laws, each of them representing the behavior with the observed frequency of an emission mechanism: 

\begin{equation}
   S_\nu^{\rm total} = S_0^{\rm gas} \left( \frac{\nu}{\rm 15~GHz} \right)^{\alpha_{\rm gas}} +  S_0^{\rm dust} \left( \frac{\nu}{\rm 225~GHz} \right)^{\alpha_{\rm dust}},
\end{equation}

\noindent where $S_0^{\rm gas}$ is the flux density contribution at a wavelength of 2 cm (frequency of 15 GHz), $S_0^{\rm dust}$ is the flux density contribution at a wavelength of 1.3 mm (frequency of 225 GHz), $\alpha_{\rm gas}$ and $\alpha_{\rm dust}$ are the spectral indices of the {ionized gas (free-free or synchrotron)} and dust contributions, respectively. For the fitting, we impose some constraints, namely that $S_0^{\rm gas}$ and $S_0^{\rm dust}$ cannot be larger than the measured total flux densities at 2 cm and 1.3 mm, respectively, that $\alpha_{\rm gas}$ should be within the range \mbox{[-1.0 , 2.0]}, and that $\alpha_{\rm dust}$ should be within the range [2.0, 4.0]. The space parameter is explored by using the Markov Chain Monte Carlo (MCMC) algorithm {with 20,000 iterations and 40 walkers. The initial values of each walker is set randomly (with a flat probability) within the allowed range for each parameter. For the final statistics, we only consider the last 10,000 iterations}. The SED fits are shown in Figs.\,\ref{fig:sed_illustrative} (one illustrative case) and \ref{fig:all_seds} (all cases). 

The illustrative SED of DR Tau shown in Fig.\,\ref{fig:sed_illustrative} reveals that some non dust emission is detectable and starts to dominate over the dust emission at wavelengths longer than 1 cm. For this specific case, our model determines a non dust emission at 2 cm of 0.10$^{+0.03}_{-0.06}$ mJy and a spectral slope $\alpha_{\rm gas}$ of 1.1$^{+0.7}_{-1.4}$. The $\alpha_{\rm gas}$ constrained for the sample spans from 0.3 to 1.1 (with 15 of 21 being between 0.5 and 1.0). Since these are the slopes expected from free-free emission \citep{Panagia1975, Reynolds1986}, hereafter we refer to the ionized gas emission as free-free emission. All fluxes and slopes constrained by our model are listed in Table  \ref{table:sed_fit}.

\subsection{Free-free emission} \label{sec:free-free}
All SEDs are well fitted by our model from Sect.\,\ref{sec:sed}, supporting a view where the extra emission at radio wavelengths is free-free emission. Overall, we determined the presence of some free-free emission at 2 cm in all sources. In absolute terms, our measured free-free emission spans from the barely detectable value of 0.02 mJy (CIDA 9 and V409 Tau) to the very large value of the Herbig AeBe star MWC 480 (0.24 mJy) and that of the spectroscopic binary DQ Tau (0.26 mJy). The latter source is the only case where the SED fitting presents some criticalities that are discussed in Appendix \ref{appendix:setup}. Furthermore, the peculiar case of HP Tau with very poor photometry (see Fig.\,\ref{fig:all_seds}) is not treated by our model. For this source, the photometric value is high enough (7 mJy) to assure that any dust contribution is negligible (since a simple fit to the available ALMA photometry results in a dust contribution at 2 cm of less than 0.1 mJy). 

In relative terms, our model constrains that the free-free emission contributes on average 65\% of the total flux at 2 cm. The scatter of this fraction across the sample is very narrow ($\sigma=8\%$) meaning that, for Class II objects at 2 cm, an equal partition between free-free and dust emission can be coarsely assumed. In contrast, the contribution to the total flux at 1 cm by the free-free emission spans from 10\% (DL Tau) to 75\% (GK Tau), with an average of 35\%, meaning that each case should be evaluated individually. At 3 mm, the distribution of this fraction is again quite narrow around a median value of 5\% while it is less than 2\% in most cases at 1.3 mm (see histogram in Fig.\,\ref{fig:ff_fraction}).     

The origin of the free-free emission from a relatively large sample can be investigated from its relation with other observable quantities. First, our estimates of the free-free emission scaled at a same distance has a loose relation with the stellar mass of Table \ref{table:properties} (Pearson coefficient $r=0.51\pm 0.19$) although the narrow interval of stellar masses involved (more than a half of the sample between 0.4 and 0.6 M$_\odot$) may conceal any tighter connection. Inspired by the correlation between the accretion rate and the free-free emission that was recently reported by \citet{Rota2024} for transition disks, we relate these two quantities in Fig.\,\ref{fig:acc-ff}. Our estimates of the free-free emission come with relatively large error bars owing to the unresolved nature of the 2-cm emission and the consequent need for the model of Sect.\,\ref{sec:sed}. Nonetheless, a clear correlation (with $r=0.66 \pm 0.17$) between this emission and the mass accretion from Table \ref{table:properties} is visible from Fig.\,\ref{fig:acc-ff}. Low accretors (in the framework of this sample, sources with less than $10^{-8}$ M$_\odot$ yr$^{-1}$) show a free-free emission at 2 cm and at 140 pc of less than 0.1 mJy, while the opposite is shown by high accretors. The correlation holds if we instead consider the total flux observed at 2 cm (without employing any model to determine the free-free emission). This is not surprising if we consider the aforementioned fraction of free-free over total flux at 2 cm that is rather constant across the sample.

Interestingly, the trend of Fig.\,\ref{fig:acc-ff} does not reveal any segregation between the sub-structured disks and the compact disks by \citet{Long2019} (see Sect.\,\ref{sec:targets}) with both categories being equally distributed across the observed interval of values (resulting in a Pearson $r$ between free-free emission and disk extent from Table \ref{table:properties} as low as 0.06). This would intuitively indicate that the origin of the correlation is completely independent of the presence of disk sub-structures at more than 50 au from the star.

\citet{Rota2024} concluded that the free-free emission determined for their sample is related to the ionized jet and that the observed trend reflects the connection between the outflow and accretion activities.
Further analyses on the free-free emission of our sample and an in-depth discussion of the implication of Fig.\,\ref{fig:acc-ff} are given in a forthcoming publication by Rota et al.\,in prep.

\subsection{Dust spectral indices} \label{sec:dust_indices}
The slope of the millimeter SED $\alpha$ defined as
\begin{equation}
\alpha=\frac{d \log{F_\nu}}{d \log{\nu}}    
\end{equation}
is a useful diagnostic of the dust properties since, for optically thin emission in the Rayleigh-Jeans regime, it relates to the dust opacity spectral index $\beta$ through $\beta=\alpha-2$. However, an important restriction on this conversion is imposed by the aforementioned optically thick nature of a significant extent of disks up to 3 mm \citep[e.g.,][]{Ribas2020, Sierra2020, Tazzari2021a, Xin2023}. Therefore, the measurement of $\alpha$ at \mbox{(sub-)centimeter} wavelengths represents a viable tool to alleviate this problem since we can assume that this emission is genuinely optically thin.

The 0.89-1.33mm spectral index $\alpha_{\rm 1mm}$ extracted from \citet{Andrews2013} for our sources is on average 2.4 (with a $\sigma=0.3$). Rather lower values (2.0) are found from the recent SMA survey of Taurus by \citet{Chung2024}, although the difference is more marginal within the intersecting sample (2.3 vs 2.1). Similarly, the 1.33-3mm index $\alpha_{\rm 3mm}$ of our sample from the photometry introduced in Sect.\,\ref{sec:sed} is on average 2.3 (with a $\sigma=0.4$). A possible interpretation of integrated values so close to 2 is that the optically thick portion of the disk at these wavelengths is lowering the actual averaged $\alpha$ of the disk. We therefore make use of our centimeter photometry to extend these indices to longer wavelengths. 

\begin{figure}
   \includegraphics[width=9cm]{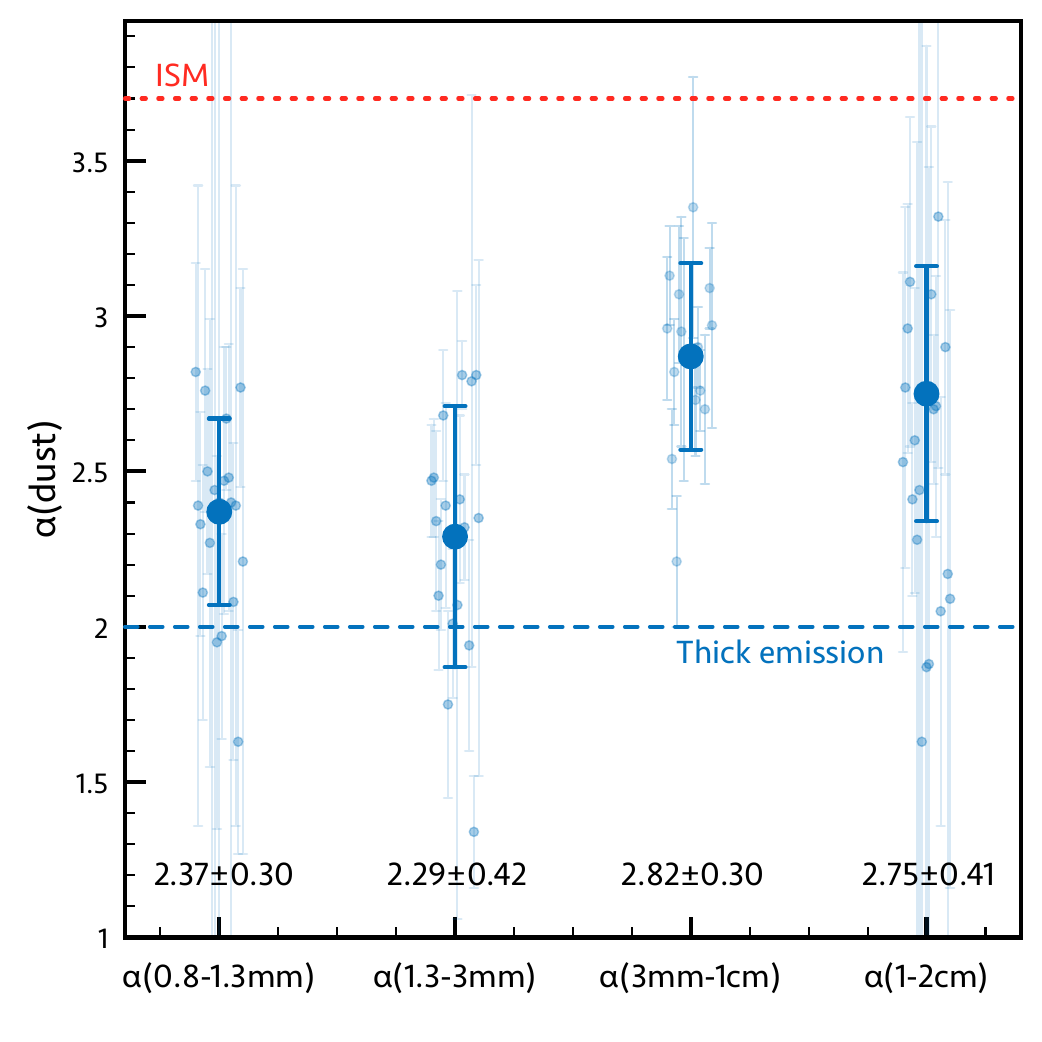}
   \caption{Millimeter and centimeter dust spectral indices. The individual $\alpha$ at four different spectral ranges are shown in semi-transparency. These are obtained removing the free-free emission from our photometry and comparing with literature values (see text). The big foreground symbols are the average values for the entire sample. The error bars of the individual sources are inherited from the uncertainty on the photometry while those of the averages are the dispersion of the relative sample. The red line indicates the spectral index of the interstellar dust grains in the millimeter while the blue line the expected index from optically thick emission.}
   \label{fig:indices}%
\end{figure}

A first hint of higher values of $\alpha$ comes from the $\alpha_{\rm dust}$ values of our model (see Sect.\,\ref{sec:sed}) shown in Table \ref{table:sed_fit}. Their average is 2.7 with a low dispersion ($\sigma=0.2$). Thus, we extract the individual indices from the observed 1-cm and 2-cm photometry after removing the free-free contribution described in Sect.\,\ref{sec:free-free}. The same procedure is applied to the millimeter photometry even though the contribution from the free-free in that regime is on average only 5\% at most. In line with the $\alpha_{\rm dust}$ from the model, the index between 3 mm and 1 cm $\alpha_{\rm 1cm}$ turned out to be 2.8$\pm$0.3 which is substantially higher than the aforementioned 2.4$\pm$0.3 from $\alpha_{\rm 1mm}$. 

Finally, we measure the index at even longer wavelengths exploiting the 1-cm and 2-cm photometry. In this regime, the impact of the free-free emission is significant (approximately 65\%, see Sect.\,\ref{sec:free-free}) and the error bars on the index are larger since the flux is detected with lower significance. Therefore these indices, in absence of any resolved image, are inevitably rather uncertain. In fact, our measured indices $\alpha_{\rm 2cm}$ between the 1-cm and 2-cm photometry turned out to be much more varied than all other indices. In particular, six sources have an error larger than 1.0 which makes the measured index meaningless. Excluding these sources from the overall measurement, the average (2.7$\pm$0.4) is very similar to that of the $\alpha_{\rm 1cm}$. Figure \ref{fig:indices} summarizes these findings manifesting that the averaged spectral indices have a pronounced change between the \mbox{(sub-)millimeter} and the \mbox{(sub-)centimeter} wavelengths while they remain stable within the two regimes.    

\begin{figure}
   \includegraphics[width=9cm]{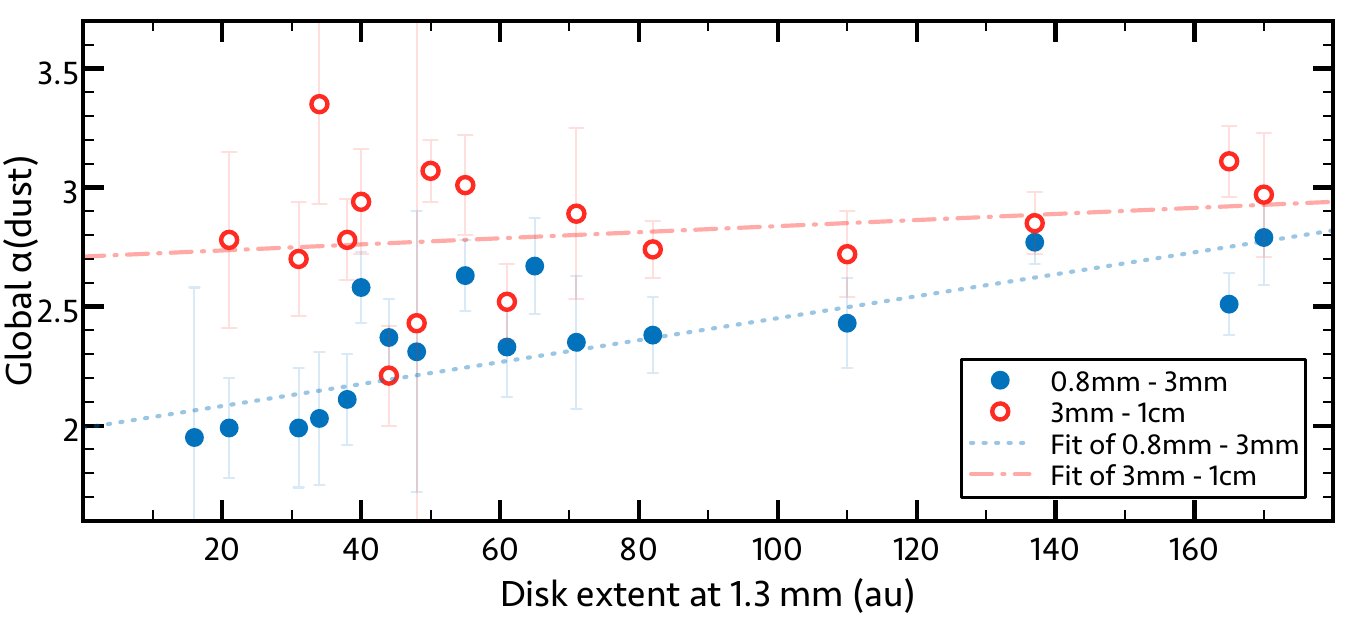}
   \caption{Dust spectral indices versus disk extent. All available $\alpha$ of our sample obtained after removing the free-free emission are compared with the disk extent from the ALMA continuum emission at 1.3 mm listed in Table \ref{table:properties}. The best fits to the data are shown as dashed lines. A small offset in extent is given to some datapoints so that indices at the same radius, when available, are from the same object.}
   \label{fig:indices_radius}%
\end{figure}

A further step to evaluate the measured spectral indices is the comparison with the disk extent measured from ALMA \citep[see Table \ref{table:properties}]{Long2019}. For this exercise, we use the $\alpha$ between 0.89 mm and 3 mm since the intermediate indices within this interval are very similar (see Fig.\,\ref{fig:indices}) and between 3 mm and 1 cm since these have smaller errors than those between 1 cm and 2 cm. As is clear from Fig.\,\ref{fig:indices_radius}, the five smaller disks in the sample ($r$ < 40 au) exhibit a millimeter index well consistent with 2.0 while larger disks have a significantly larger value (on average 2.4). A similar trend of larger indices for larger disk extents is found by \citet{Tazzari2021a} for $\alpha$ at the same wavelengths and by \citet{Chung2024} for $\alpha$ between 0.85 and 1.3 mm. Conversely, our measured $\alpha$ between 3 mm and 1 cm is significantly larger for smaller disks while is only mildly larger for larger disks. This trend results in an index distribution with the disk extent that is nearly constant. As is discussed in Sect.\,\ref{sec:discussion_cm}, the behavior described in this section is consistent with a view where the inner few tens of au of disks are optically thick up to $\sim$3 mm while the optical depth severely decreases at longer wavelengths. It also highlights that the sub-centimeter indices from the optically thin emission of sub-structured and compact disks are not significantly different (see the flat trend in Fig.\,\ref{fig:indices_radius}).

\subsection{Dust integrated emission} \label{sec:dust_masses}
The dust emission at millimeter and centimeter wavelengths also offers the possibility to probe the mass budget in solids. For optically thin emission in the Rayleigh-Jeans regime, the integrated flux is in fact directly converted into dust mass under some assumptions on the dust opacity (discussed in Sect.\,\ref{sec:discussion_uncertainties}). Here we compare the millimeter and centimeter fluxes of sub-structured and compact disks (Sect.\,\ref{sec:targets}) of the sample leaving aside the conversion to the actual mass.  

\begin{figure}
   \includegraphics[width=9cm]{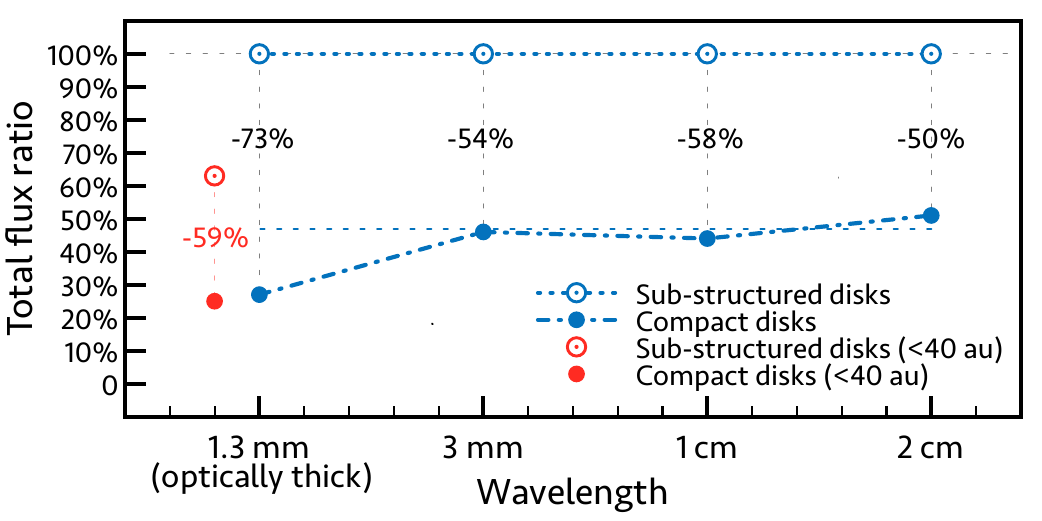}
   \caption{Total flux ratio between sub-structured and compact disks at different wavelengths. The average dust integrated flux at a same distance of compact disks from 1.3-mm, 3-mm, 1-cm, and 2-cm fluxes is confronted with that of sub-structured disks. Fluxes at 7 mm are not shown because the sub-samples are too small. The red points indicate the fluxes integrated over the inner 40 au from the ALMA highly resolved images. At 1.3 mm, the compact disks have relatively lower fluxes than at the other wavelengths while a similar value (50\%--60\%) is found when considering only the inner 40 au, suggesting that the emission from this disk region is severely optically thick.}
   \label{fig:mass_ratio}%
\end{figure}

As is clear from Fig.\,\ref{fig:mass_ratio}, the average total flux of sub-structured disks\footnote{Here we use a disk extent of 50 au to discriminate between the two categories. This criterion is nearly identical to the sub-structured vs compact disk classification except for the massive but formally compact disk of DR Tau (see Sect.\,\ref{sec:targets}). Also, DQ Tau has been removed from this analysis in view of the large uncertainty at centimeter wavelengths that leads to anomalously high fluxes (see Sect.\,\ref{sec:discussion_cases} and Appendix \ref{appendix:setup}).} normalized at a same distance is 73\% larger than that of compact disks. However, this value is decreased to 54\% at 3 mm and then remains constant up to 2 cm. It is thus tempting to conclude that we are seeing another manifestation of the optically thick nature of the 1.3-mm emission. In fact, the mass of compact disks would be more under-estimated than that of extended disks because of having a larger portion of optically thick emission \citep[see also][]{Sanchez2024}.  

To test this interpretation, we repeat the same exercise with the highly resolved flux from the 1.3-mm ALMA images. As expected, diminishing the aperture over which the total flux is integrated results in lowering the difference between sub-structured and compact disks. In particular, Fig.\,\ref{fig:indices_radius} had suggested that the inner 40 au of disks could be heavily optically thick. Interestingly, when we consider an aperture of 40 au (which is still larger than the extent of most compact disks) then sub-structured disks are only 59\% brighter than compact disks (see red points in Fig.\,\ref{fig:mass_ratio}) that is the much closer to the value found on the total flux at longer wavelengths. This suggests that the compact disks of this sample have intrinsically 50\%--60\% lower masses than the extended disks and that the 73\% recorded at 1.3 mm is the effect of the optical depth being relatively more important in compact disks. This does indicate what is net loss of 1.3-mm flux in these disks but it does qualitatively suggest that is very high even for extended disks since as much as an average 65\% of the flux from this type of disks from our sample is from the inner 40 au.

\section{Discussion} \label{sec:discussion}

The tremendous progress achieved over the last decade in the characterization and understanding of the (sub-)millimeter emission from planet-forming disks is not accompanied by equally important advancement in the study of the (sub-)centimeter emission. While several studies on individual disks have been carried out \citep[e.g.,][]{Lommen2009, Perez2015, Wright2015, Macias2018, Carrasco-Gonzalez2019, Guidi2022, Curone2023}, only few authors have focused on large samples aimed at determining the general behavior of more common objects \citep{Ubach2012, Ubach2017, Norfolk2021}. This work is a further effort in this direction that exploits the potential of VLA to characterize the 12--50 GHz emission from the universe.

\subsection{The benefits of the centimeter emission from Class II} \label{sec:discussion_cm}

Notwithstanding the limited resolving power and uv coverage of currently available interferometers, the (sub-)centimeter emission from Class II objects offers several opportunities that are emphasized in this work. First of all, this type of emission from planet-forming disks is optically thin. Conversely, increasing evidence suggests that the emission from a large portion of disks is optically thick for most of the ALMA wavelengths ($\lesssim$3 mm). This work supports this idea by showing the pronounced change of the unresolved spectral index between the millimeter and the sub-centimeter regime, and that this change is remarkable in small disks where most of the disk extent has a high optical depth. In particular, our study suggests that the disk region with optically thick millimeter emission could be as large as 40 au for most disks. This finding also explains why the millimeter spectral index of disks with resolved cavity is larger than that of full disks \citep{Pinilla2014}. The most immediate benefit of the optically thin emission from $\lambda>$ 3 mm is to remove at least the optical depth among the uncertainties on the determination of the actual dust mass in the disk (see Sect.\,\ref{sec:discussion_uncertainties}) in a framework where insufficient solid mass seems to be available for the planet formation \citep[e.g.,][]{Manara2018, Mulders2021}.

Secondly, the centimeter emission provides access to larger dust grains (in principle up to 10 times larger than those probed by the millimeter emission) that are fundamental to study the grain growth \citep[e.g.,][]{Testi2014} and pressure traps \citep[e.g.,][]{Pinilla2012b} in disks. The precise determination of vertical and radial distribution of possible cm-sized dust grains is entrusted to future dedicated modeling and radio instruments with increasing resolving power. In fact, the best resolution achieved by VLA at 2 cm for an object at 150 pc is approximately 30 au, which is the whole disk extent in many cases. In this work, we also demonstrate that the possible drawback of the centimeter emission represented by the decreasing brightness of disks at these wavelengths is inconsiderable. In fact, according to our SED fitting a substantial fraction of the 2-cm emission is still originated by the dust, meaning that some tens of $\mu$Jy are expected from a typical Taurus disk. These fluxes are easily detectable with VLA in a reasonable telescope time, and will be certainly within reach of ngVLA and SKA-mid at its higher-frequency regime.  

A third advantage of studying the centimeter emission from Class II is the possibility to probe simultaneously two completely independent components like the dust and the coronal or accretion processes, with the only (current) limitation that a fit applied to multi-wavelength photometry is needed to disentangle the two types of emission. This work confirms that the free-free emission may be very common from Class II objects \citep[see e.g.][]{Rodriguez1999, Dzib2013} and provide a possible benchmark to compare with other star-forming regions \citep[e.g., Corona Australis,][]{Galvan2014}. It also contributes to elucidating that several processes are connected in the disk inner regions \citep[e.g., the free-free emission is directly related with the accretion,][which in turn is known to relate with the stellar and disk mass]{Rota2024} while a connection with peripheral disk structures (>50 au) appears much more marginal.

\subsection{The uncertainty on the centimeter emission from Class II} \label{sec:discussion_uncertainties}

Inevitably, the analysis of the centimeter emission also introduces a number of uncertainties that are outlined here. First, the entanglement of very different types of emission (e.g., dust thermal, free-free, stellar activity) from the same unresolved emission requires a model at different levels of complexity. While our SED fitting of Sect.\,\ref{sec:sed} seems to overall give good results in most cases (but see DQ Tau in Sect.\,\ref{sec:discussion_cases} and Appendix \ref{appendix:setup}), it does not foresee any variation in the slope of either the free-free or the dust emission across the spectrum. If the free-free emission may correctly be fitted with a constant slope, some curvatures in the slope of the dust emission are possible. Beside the aforementioned steepening at transition between optically thick and thin regimes, we may also expect a further steepening at centimeter wavelengths due to the dust properties or because the dust becomes cold enough to depart from the Rayleigh–Jeans regime \citep[see e.g.,][]{Ricci2017}.

Our SED fitting (see Sect.\,\ref{sec:sed}) only includes one slope for the dust emission and for the free-free emission. However, all the fluxes used to extract the dust contribution have been taken from the observations after subtracting the modeled free-free emission. Therefore, the impact of local bends in the SED is limited since the free-free emission is expected to be rather flat across the centimeter wavelengths. Nonetheless, it is still possible that our procedure tends to underestimate the free-free emission and overestimate the dust emission between 1 and 2 cm if the intrinsic dust spectrum is steepening in this regime. Again, this issue shall be solved with resolved centimeter observations disentangling the various sources of emission. 

Secondly, the study of the centimeter emission also presents critical measurements that go beyond the unresolved nature of the observed emission. The fraction of this emission that originates beyond the dust thermal processes (e.g., free-free and synchrotron) is in fact known to be highly variable over few weeks \citep{Ubach2017} or even few days \citep[see Sect.\,\ref{sec:discussion_cases}]{Salter2010}. Therefore, an accurate evaluation of these processes require monitoring surveys with different temporal baselines (days to years) that are very consuming in resources and time. 

Finally, the mass is typically considered the most critical and yet fundamental measurement of planet-forming disks. The high optical depth of the millimeter emission, although not the only one, can be the origin of notable inaccuracy in this determination \citep[see e.g.][]{Miotello2023}. A simplified approach to this goal is in fact the direct conversion of the total observed flux at a given frequency $F_\nu$ in the assumption of optically thin emission through:
\begin{equation} \label{eq:disk_mass}
    M_{\rm dust}=\frac{F_\nu d^2}{\kappa_\nu B_\nu(T_{\rm dust})}
\end{equation}
where $\kappa_\nu$ is the dust absorption opacity at that frequency and $B_\nu(T_{\rm dust})$ the Planck function at a given dust temperature $T_{\rm dust}$. With $F_\nu$ measured at wavelengths longer than 3 mm, the opacity $\kappa_\nu$ becomes decidedly the main origin of uncertainty.

Several \mbox{(sub-)millimeter} surveys \citep[e.g.,][]{Andrews2005, Pascucci2016, Cieza2019} have been used to constrain the dust mass from the total integrated flux. In absence of dedicated modeling, the common practice is to assume that the opacity $\kappa_\nu$ scales with the frequency as $\nu^\beta$ \citep{Beckwith1990}. However, different values of $\beta$ have been used in the literature spanning from 0.4 \citep[e.g.,][]{Carpenter2014} to 1.0 \citep[e.g.,][]{Ansdell2016}. Instead, \citet{Tazzari2021b} opted for $\beta$ directly measured from the data as $\alpha_{\rm 0.89-3\,mm}-2$. As for the disk temperature $T_{\rm dust}$, most authors took a constant 20 K assuming that most millimeter emission comes from the cold, isothermal, outer disk midplane. However, a dependency for $T_{\rm dust}$ on the stellar luminosities have been considered in some cases \citep[see][]{Andrews2013}, and values higher than 20 K can be expected from compact disks even around low-luminosity stars \citep{vanderMarel2023b}. In any case, a more thorough determination of the disk mass requires dedicated modeling of highly resolved observations as in e.g., \citet{Carrasco-Gonzalez2019}, \citet{Macias2021}, \citet{Guidi2022}, \citet{GuerraAlvarado2024}. 

While a thoroughly refined measurement of the disk masses based on our unresolved photometry is an unattainable task, some useful considerations can be drawn by comparing the measured disk masses at different wavelengths and with different dust opacities. To do it, the whole available photometry (see Sect.\,\ref{sec:overview}) corrected for the free-free emission (Sect.\,\ref{sec:free-free}) is used to extract the disk mass of the entire sample at different frequencies using Eq.\,\ref{eq:disk_mass}. We adopt a $T_{\rm dust}$ of 20 K for all sources to avoid introducing further dependencies. We take the aforementioned dust opacity by \citet{Beckwith1990} $\kappa_\nu=3.37\,(\nu/337\,{\rm GHz})^\beta\,{\rm cm^2\,g^{-1}}$ with different $\beta$ as well as some illustrative opacities from the literature \citep[with $a_{\rm max}=1$ mm, see their comparison in Figs.\,6 and 11 from \citealp{Birnstiel2018}]{Zubko1996, D'Alessio2001, Birnstiel2018}.

\begin{figure}
   \includegraphics[width=9cm]{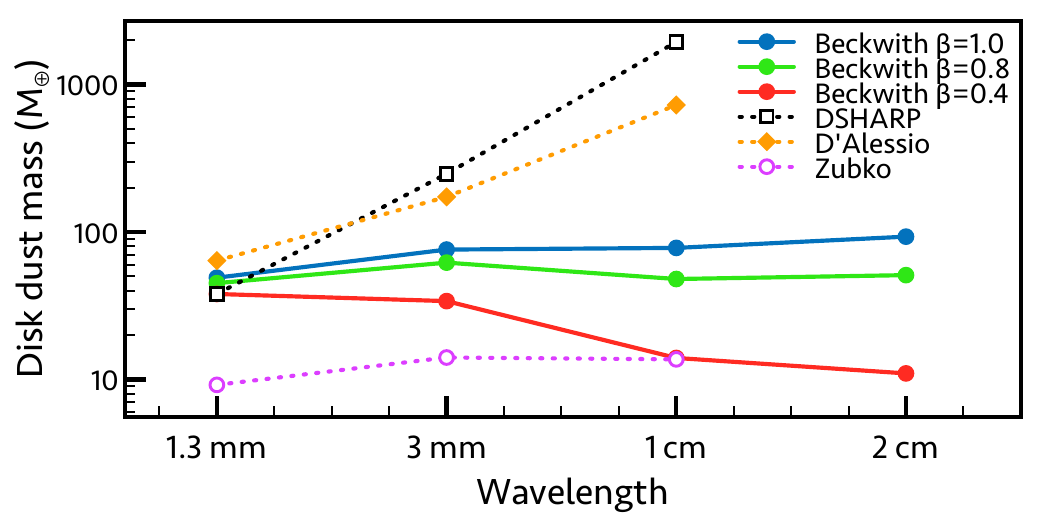}
   \caption{Average dust mass at different wavelengths and with different opacities. The average disk dust mass from our sample is calculated from 1.3-mm, 3-mm, 1-cm, and 2-cm fluxes assuming optically thin emission and with different sets of opacity \citep[with $a_{\rm max}=1$ mm]{Beckwith1990, Zubko1996, D'Alessio2001, Birnstiel2018}. A slightly different number of targets contribute to the individual datapoints across wavelengths depending on the available photometry.}
   \label{fig:mass_opacity}%
\end{figure}

The result of this exercise is shown in Fig.\,\ref{fig:mass_opacity}. By construction of the $\kappa_\nu$ formula by \citet{Beckwith1990}, the impact of the choice of $\beta$ is limited at millimeter wavelengths but is substantial at centimeter wavelengths. Using the millimeter spectral slope of our sample (see Sect.\,\ref{sec:dust_indices}) like in \citet{Tazzari2021b} would mean taking an average $\beta$ of 0.4 \citep[like in e.g.,][]{Carpenter2014, Pascucci2016}. Instead, taking the $\alpha$ obtained at longer wavelengths where the contamination from optically thick emission is minimized would lead to the adoption of $\beta$=0.8 that, by construction of Eq.\,\ref{eq:disk_mass}, leads to the same value of masses calculated at all wavelengths. The $\beta$=1 that is sometimes used in previous work clearly leads to similar results. Conversely, the other sets of opacities reveal different behaviours. On the one hand, those by \citet{D'Alessio2001} and \citet[the DSHARP]{Birnstiel2018} are very similar to our nominal attempt at millimeter wavelengths while their very low value at centimeter wavelengths yields possibly unrealistic values of masses constrained from centimeter emission. On the other hand, the opacity by \citet{Zubko1996} leads to very low masses across the entire available spectrum only approaching those by \citet{Beckwith1990} in the case of $\beta=0.4$. Somehow similar results are found when looking at opacities with $a_{\rm max}=1$ cm instead of 1 mm. While inconclusive for a refinement of the disk mass, Fig.\,\ref{fig:mass_opacity} emphasises that, in the assumption of maximum grain sizes of at least 1 mm (which is justified by the strong centimeter emission detected), the uncertainties on the dust opacity for the centimeter emission are very large, and even larger than for the millimeter emission.

\subsection{The notable sources in the sample} \label{sec:discussion_cases}
By construction of the sample, only ordinary objects are treated in this work. As outlined in Sect.\,\ref{sec:targets}, none of these disks looks extremely extended, bright, nor asymmetric like e.g., AB Aur, HD142527, or IRS 48 \citep{Boccaletti2020, Casassus2012, vanderMarel2013}. Instead, the absence of any large cavity in these disks results in the outer disk being self-shadowed and thus faint in scattered light \citep{Garufi2017, Garufi2022b} with direct implication on the disk surface density and flaring angle. Nonetheless, the centimeter view of these ordinary objects highlights a few notable cases that are discussed here.  

CIDA 9 and IP Tau are the only sources with an evident disk cavity from the ALMA images by \citet{Long2019}. Their free-free emission is low, the second and fifth weakest in the sample. Another source with similarly low free-free emission (third weaker) is DS Tau, hosting a disk where a bright ring accounts for $\sim$80\% of the total ALMA flux. The presence of a central flux component detected by ALMA is the only element that marks a difference from the disk of CIDA 9 and IP Tau. It is therefore tempting to look at the other two of the five sources with low free-free emission. These are V409 Tau and GI Tau, which both host a full, compact disk when imaged by ALMA. However, both disks are relatively bright in scattered light given their very low dust mass \citep[similarly to IP Tau, see][]{Garufi2024}, and this morphology may be suggestive of a small-scale cavity that is not resolved by ALMA (that in the case of V409 Tau is also suggested by a small near-IR excess). In fact, recent high-resolution ALMA images of GI Tau reveal a deep disk gap with a central, unresolved component like DS Tau (Long et al.\,in prep.). Therefore, while overall there is no obvious connection between the outer disk morphology and the free-free emission (see Sect.\,\ref{sec:free-free}), the considerations on these five objects seem to highlight a possibly decreased accretion/ejection processes in the presence of a (small-scale) disk inner cavity, as also seen from disks with more prominent disk cavities \citep{Najita2007, Najita2015, Manara2016}.

An additional sub-category with peculiar characteristics is that of the massive though compact disks. DO Tau, DQ Tau, and DR Tau all host a disk with 50 au or less in size (lying in the lowest half of the distribution) and with 70 M$_\oplus$ or more in dust mass (being among the top five). These sources exhibit three of the four highest free-free emissions and three of the five highest accretion rates of the whole sample. More interestingly, DO Tau and DR Tau are two of the three sources showing evidence of ambient emission from SPHERE (with the third being HP Tau, see below). As shown by \citet{Huang2022} and \citet{Winter2018}, DO Tau shows evidence of extended outflow processes and interaction with the environment as well as the nearby star HV Tau. On the other hand, DR Tau shows extended arc-like structures that are reminiscent of an encounter with a cloudlet \citep{Mesa2022}. This may speculatively suggest that the disk of these objects accrete from the environment, and that this accreted material may in turn boost the stellar accretion and outflow processes. 

On the contrary, the spectroscopic binary DQ Tau does not show any ambient material from the SPHERE image \citep{Garufi2024}. However, this source is known to exhibit strong flares at millimeter wavelengths due to the variable synchrotron emission following magnetic interaction between the individual stars \citep{Salter2010} or periodic pulsed accretion induced by streamers from the circumbinary disk \citep{Tofflemire2017}. The extremely high flux recorded in our K-band observation (see Fig.\,\ref{fig:all_seds}) may in principle indicate an episode of outburst within a normally high level of accretion activity. However, this episode should not impact the Ka-band photometry since the two observations are obtained on the same day (12 December 2021), or alternatively the event should be shorter-living than very few hours. All this said, it is not surprising that this object represents a challenge for the model (see Sect.\,\ref{sec:free-free} and Appendix \ref{appendix:setup}).   

Finally, the case of HP Tau is the most anomalous of the sample. As shown in Sect.\,\ref{sec:free-free}, the very high flux recorded in the Ku band (see Fig.\,\ref{fig:all_seds}) makes it the only outlier of the trend for the non dust emission with the accretion rate of Fig.\,\ref{fig:acc-ff}. This star is part of a small group with CoKu HP Tau G3 and V1025 Tau, and is surrounded by a bright reflection nebula. A clear interaction with the environment is visible from the SPHERE image \citep{Garufi2024}. It is tempting to ascribe the high 2-cm flux to an episode of increased accretion and outflow. However the stellar light curve only shows minor variability in the visible \citep[half a magnitude from the ASAS-SN survey,][]{Shappee2014}, and no major event is recorded before the VLA observation date (February 2020). Thus, an episodic accretion cannot explain the discrepancy between the accretion rate and the radio flux, leaving open the possibility that the latter is a different type of emission (such as synchrotron emission). Future multi-band, multi-epoch radio observations are therefore needed to clarify this case.

\section{Summary} \label{sec:conclusions}
This work takes a broad view of the centimeter emission from Class II sources by analyzing a set of 21 objects from the Taurus star-forming region with VLA observations available. Ten of these stars host extended, sub-structured disks following the classification from the ALMA millimeter survey by \citet{Long2019}, while 11 host compact disks with no evidence of resolved sub-structures. The VLA datasets consist of unresolved or barely resolved images in the Q (7 mm, {9 sources available}), Ka (1 cm, {19 sources}), K (1.4 cm, {9 sources}), Ku (2 cm, {all 21 sources}), and C (6 cm, {10 sources}) bands from which we extract the photometry. Each source is detected in all bands except the C band where only two sources are detected.

We fit the SED of the entire sample constructed from the VLA observations plus some literature data spanning from 0.85 mm to 6 cm. The fit includes two power laws to account for the possible non dust emission at centimeter wavelengths. Our results can be summarized as follows.

\begin{itemize}
    \item The presence of non dust emission dominating at $\lambda$>1 cm is ubiquitously determined in the sample. A spectral slope between 0.3 and 1.1 was characterized by the model suggesting free-free emission from the disk inner regions.  
    \item The free-free emission at 2 cm scaled at a same distance correlates well with the accretion rate. This is suggestive of a strong link between the accretion and outflow activities that is investigated in a forthcoming work. Remarkably, sub-structured and compact disks are randomly distributed along the trend indicating that the presence of large-scale disk structures has no impact on the accretion and outflow closer in. Yet, a possible connection for the free-free emission with small-scale cavities may be present.
    \item As much as 35\% of the total flux at 2 cm is from the dust revealing that disks are still relatively bright at these wavelengths. With the SED fit, we can therefore obtain the centimeter dust spectral indices (disentangled from the free-free emission). We find an abrupt change in the median value from 2.4 at the ALMA frequencies to 2.8 in the VLA regime. This change likely sets the transition from optically thick emission (ALMA) to thin emission (VLA).   
    \item The distribution of the millimeter and centimeter indices with the disk radial extent by ALMA highlights that the difference between the two is more pronounced for compact disks. This supports the idea of a high optical depth for the inner 40 au up to 3 mm since a larger relative portion of the compact disk provides optically thick millimeter emission compared to a larger disk. Instead, centimeter spectral indices of the two disk categories are similar suggesting that the grain population responsible for these indices are analogous in different disk types.
    \item In principle, the disk mass in solids can be refined from the (sub-)centimeter, optically thin emission. However, we emphasize that at 1--2 cm the choice of the dust opacity plays a major role. We show that up to two orders of magnitude different estimates are found from our centimeter flux based on a set of opacities from the literature that assume millimeter maximum grain sizes, which is even larger than at millimeter wavelengths where, nevertheless, the high optical depth plays a major role. 
\end{itemize}

This work contributes to the general effort by the community to characterize less exceptional planet-forming disks with a broader, multi-wavelength approach. These VLA observations stress the usefulness of the centimeter emission to study multiple components of Class II objects and access more truthful dust mass and grain properties in their disks. While the moderate resolving power of the current generation of radio telescopes is hampering the employment of this emission, the next decade is promising its full exploitation thanks to the onset of ngVLA and SKA-mid whereas the ALMA Band 1 will already provide useful constraints on the sub-centimeter emission in the near future.

\begin{acknowledgements}
      We thank {the referee for the insightful comments that improved the manuscript}, and A.\,Banzatti and P.\,Pinilla for very helpful discussions. The National Radio Astronomy Observatory is a facility of the National Science Foundation operated under cooperative agreement by Associated Universities, Inc. This research has made use of the VizieR catalogue access tool, CDS, Strasbourg, France (DOI: 10.26093/cds/vizier). The original description of the VizieR service was published in Ochsenbein et al. (2000). This work was supported by the PRIN-INAF 2019 Planetary Systems At Early Ages (PLATEA) and by the Large Grant INAF 2022 YODA (YSOs Outflows, Disks and Accretion: towards a global framework for the evolution of planet forming systems). The interaction among authors have been appreciably fostered by the INAF mini-grant of the "Bando di finanziamento della Ricerca Fondamentale 2023". Part of the research activities described in this paper were carried out with contribution of the Next Generation EU funds within the National Recovery and Resilience Plan (PNRR), Mission 4 - Education and Research, Component 2 - From Research to Business (M4C2), Investment Line 3.1 - Strengthening and creation of Research Infrastructures, Project IR0000034 – "STILES - Strengthening the Italian Leadership in ELT and SKA". C.C.-G. acknowledges support from UNAM DGAPA-PAPIIT grant IG101224 and from CONAHCyT Ciencia de Frontera project ID 86372. C.F.M. and S.F. are funded by the European Union (ERC, WANDA, 101039452 and ERC, UNVEIL, 101076613, respectively). Views and opinions expressed are however those of the author(s) only and do not necessarily reflect those of the European Union or the European Research Council Executive Agency. Neither the European Union nor the granting authority can be held responsible for them. S.F. also acknowledges financial contribution from PRIN-MUR 2022YP5ACE. P.C. acknowledges support by the Italian Ministero dell'Istruzione, Universit\`a e Ricerca through the grant Progetti Premiali 2012 – iALMA (CUP C52I13000140001) and by the ANID BASAL project FB210003. Support for F.L. was provided by NASA through the NASA Hubble Fellowship grant \#HST-HF2-51512.001-A awarded by the Space Telescope Science Institute, which is operated by the Association of Universities for Research in Astronomy, Incorporated, under NASA contract NAS5-26555. C.J.C. acknowledges support from the Science \& Technology Facilities Council (STFC) Consolidated Grant ST/W000997/1. This work has also been supported by the European Union's Horizon 2020 research and innovation programme under the Marie Sklodowska-Curie grant agreement No 823823 (DUSTBUSTERS).
\end{acknowledgements}

\bibliographystyle{aa} 
\bibliography{Reference} 

\appendix

\section{Individual images and SEDs} \label{appendix:setup}
\begin{figure}
    \centering
    \includegraphics[width=8.5cm]{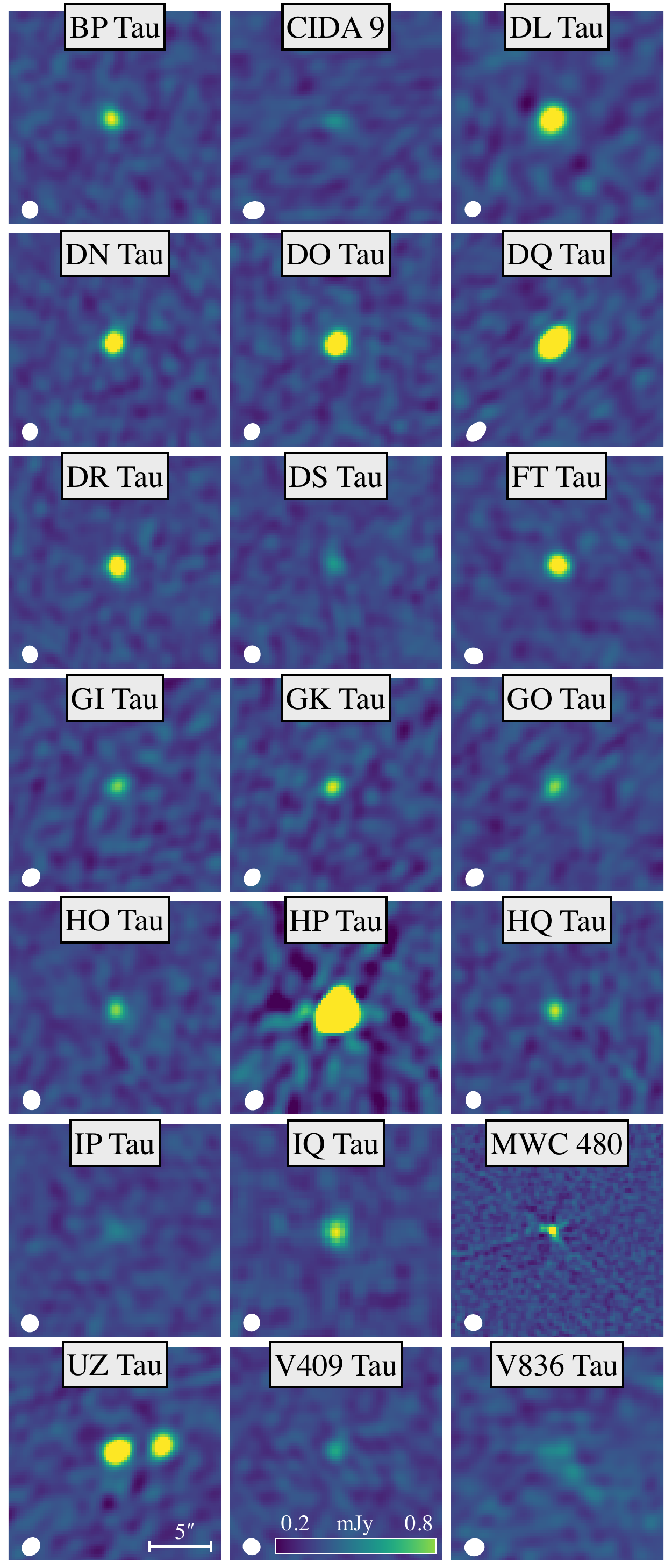}
    \caption{VLA imagery. All images in the Ku band (2 cm) are shown with the same spatial and color scale (shown to the bottom of the image). The beam size is shown to the bottom left of each panel.}
    \label{fig:images_ku}
\end{figure}

\begin{figure*}
  \centering
   \includegraphics[width=5.5cm]{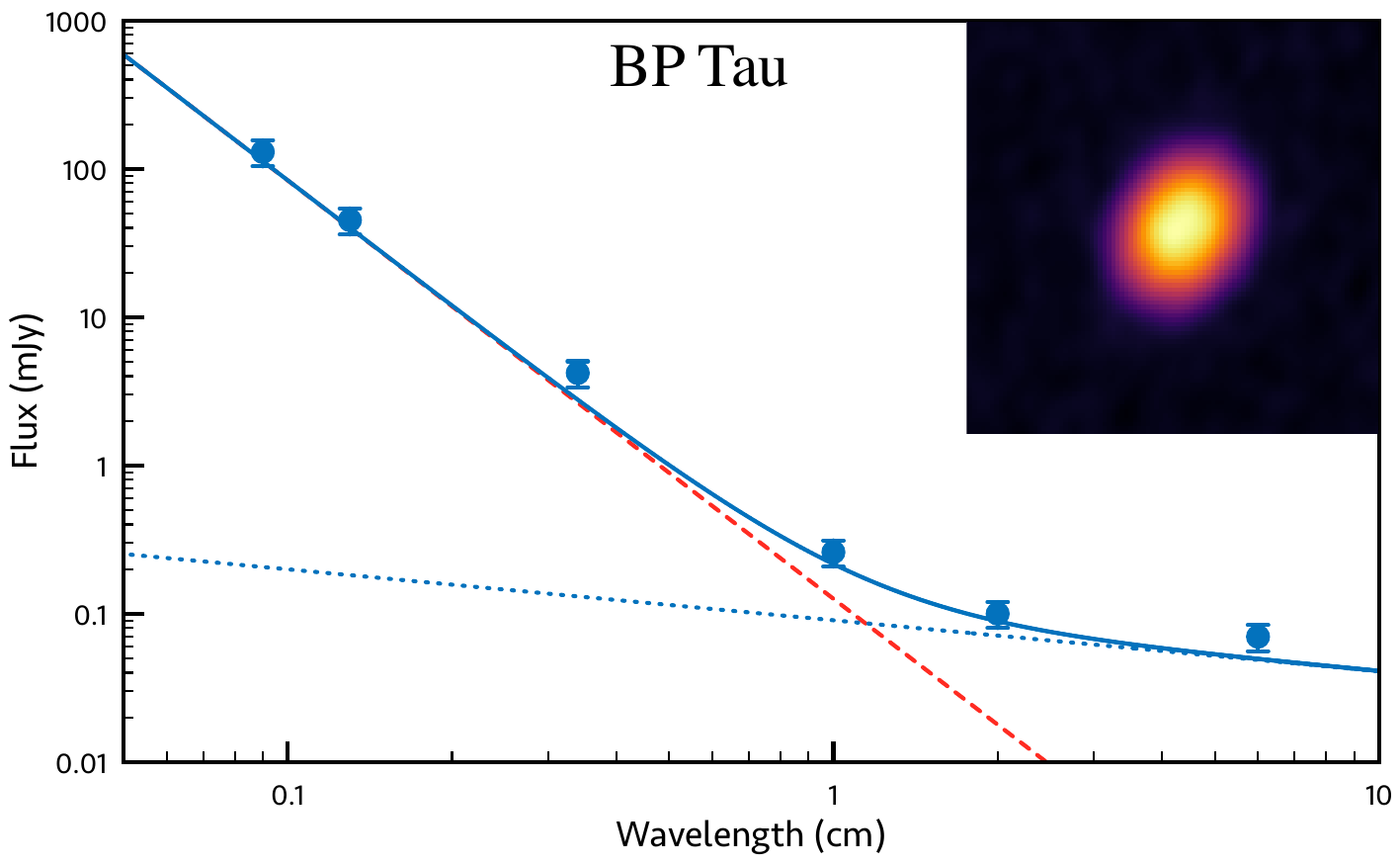}
   \includegraphics[width=5.5cm]{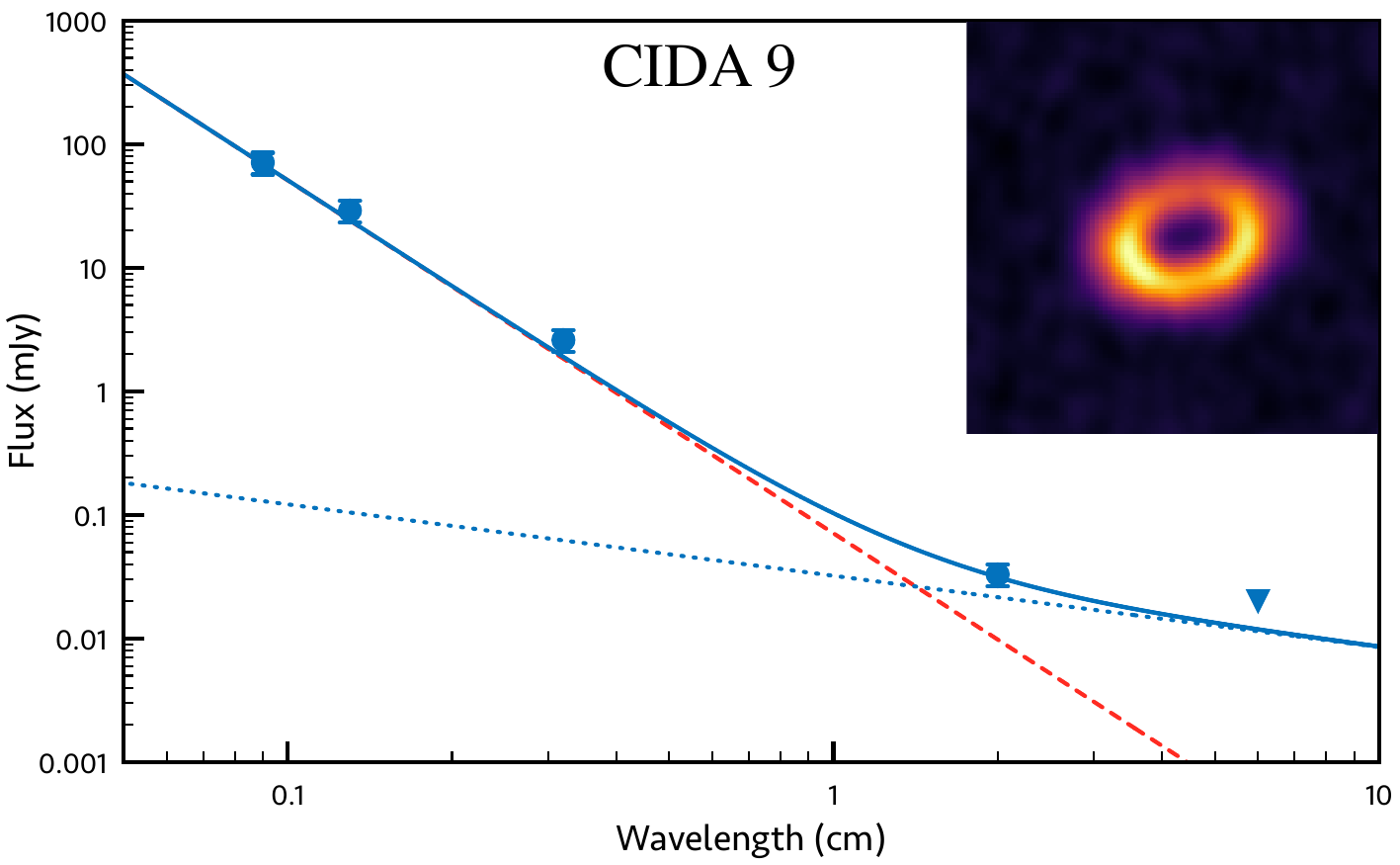}
   \includegraphics[width=5.5cm]{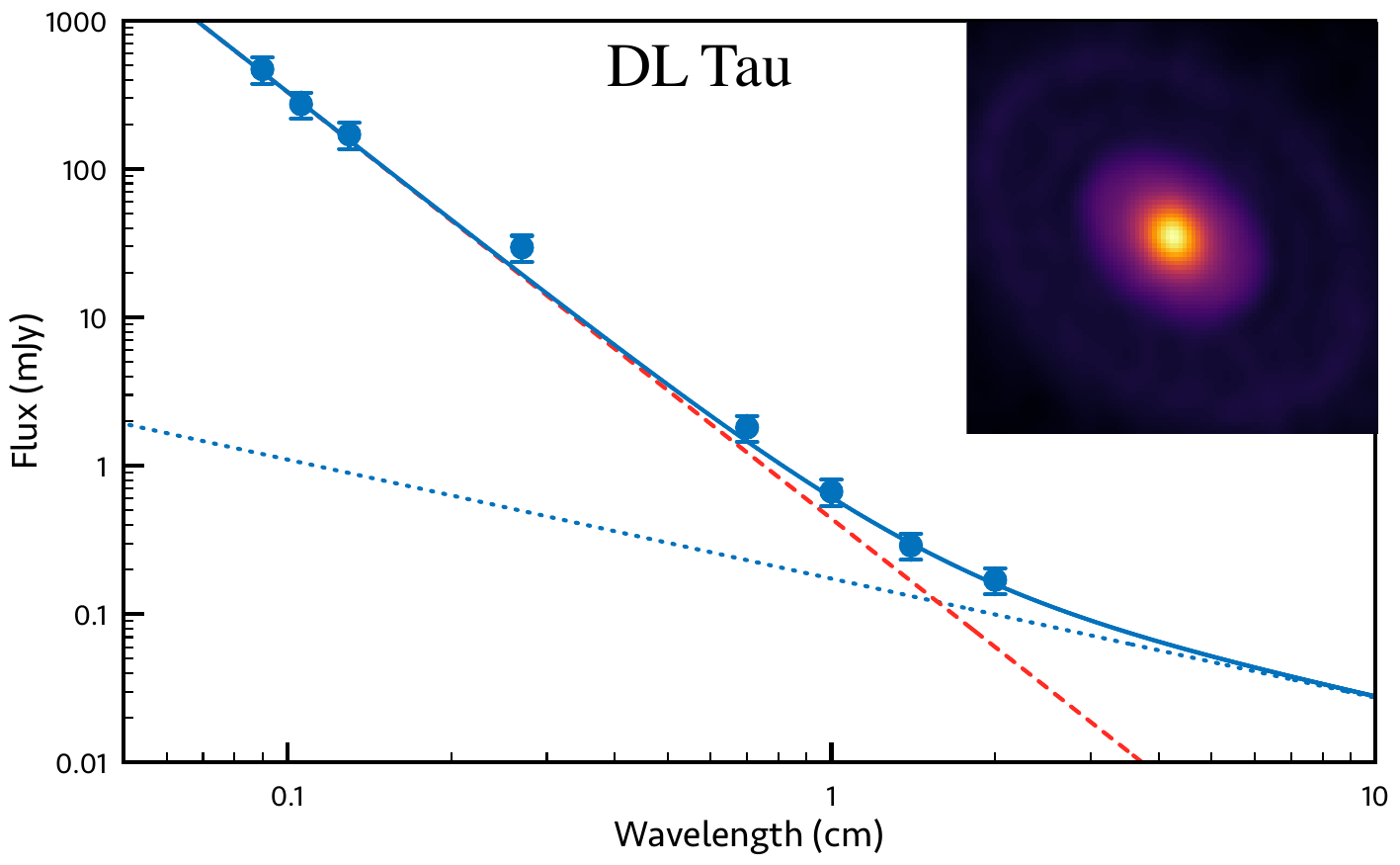}
   \includegraphics[width=5.5cm]{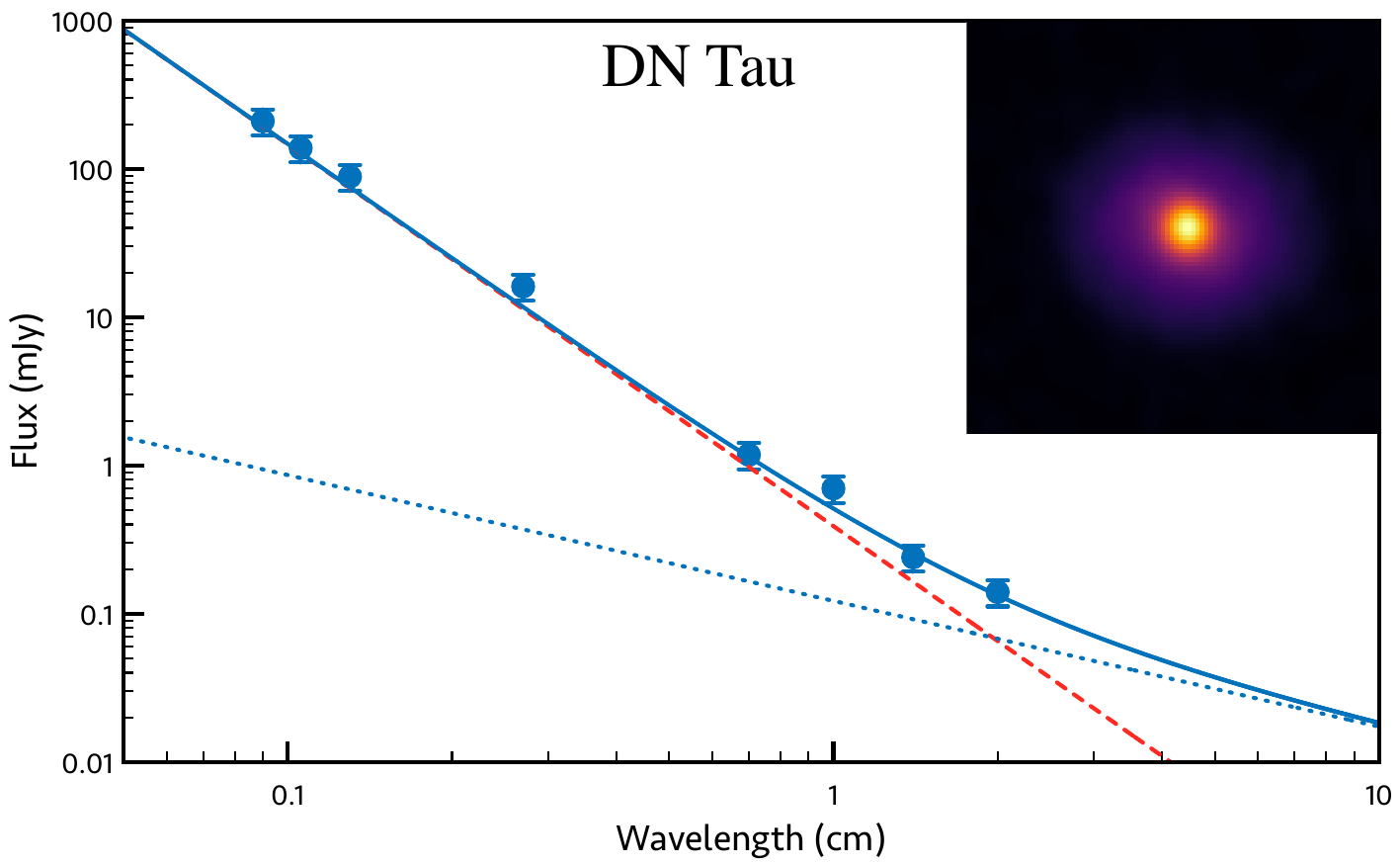}
   \includegraphics[width=5.5cm]{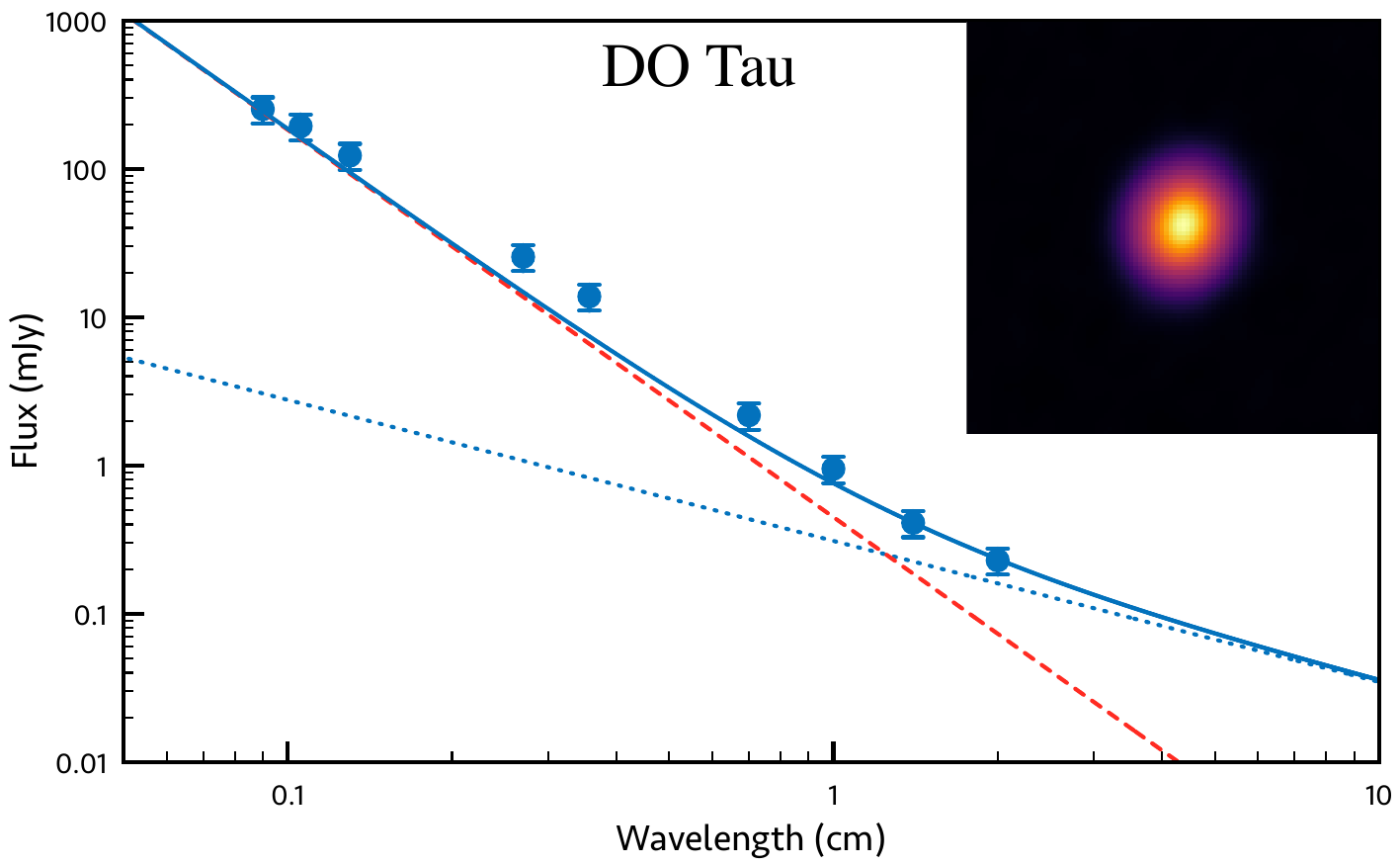}
   \includegraphics[width=5.5cm]{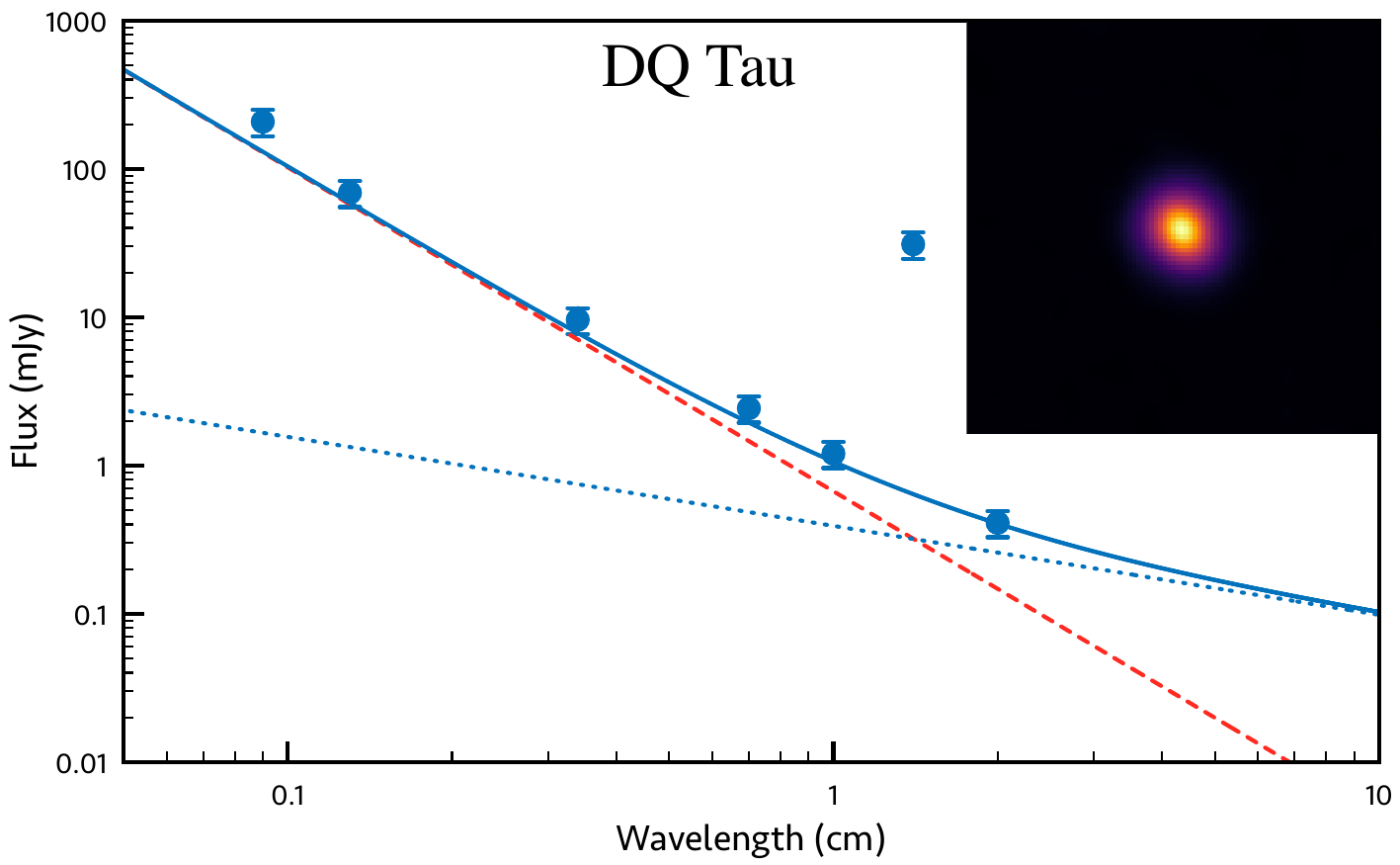}
   \includegraphics[width=5.5cm]{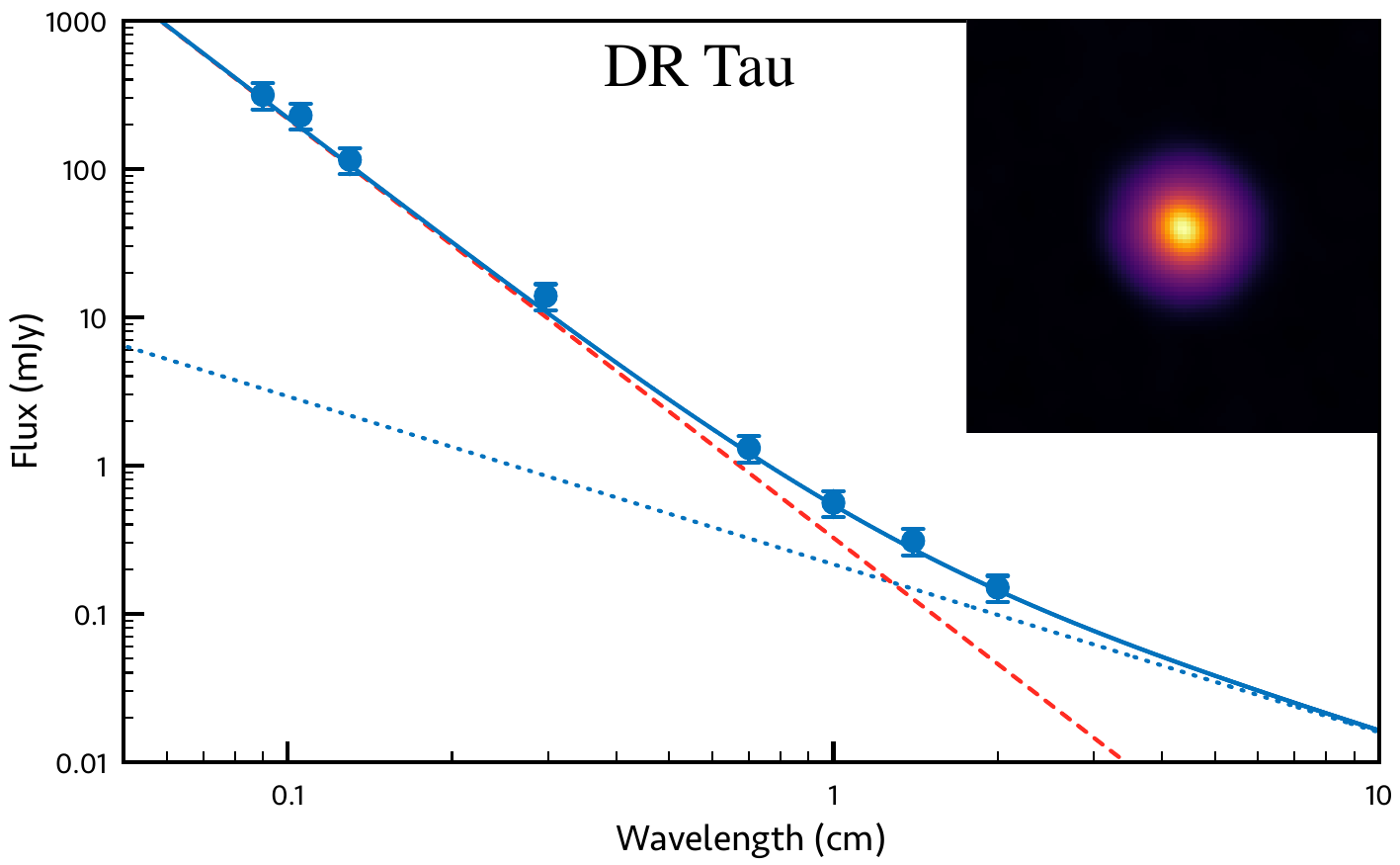}
   \includegraphics[width=5.5cm]{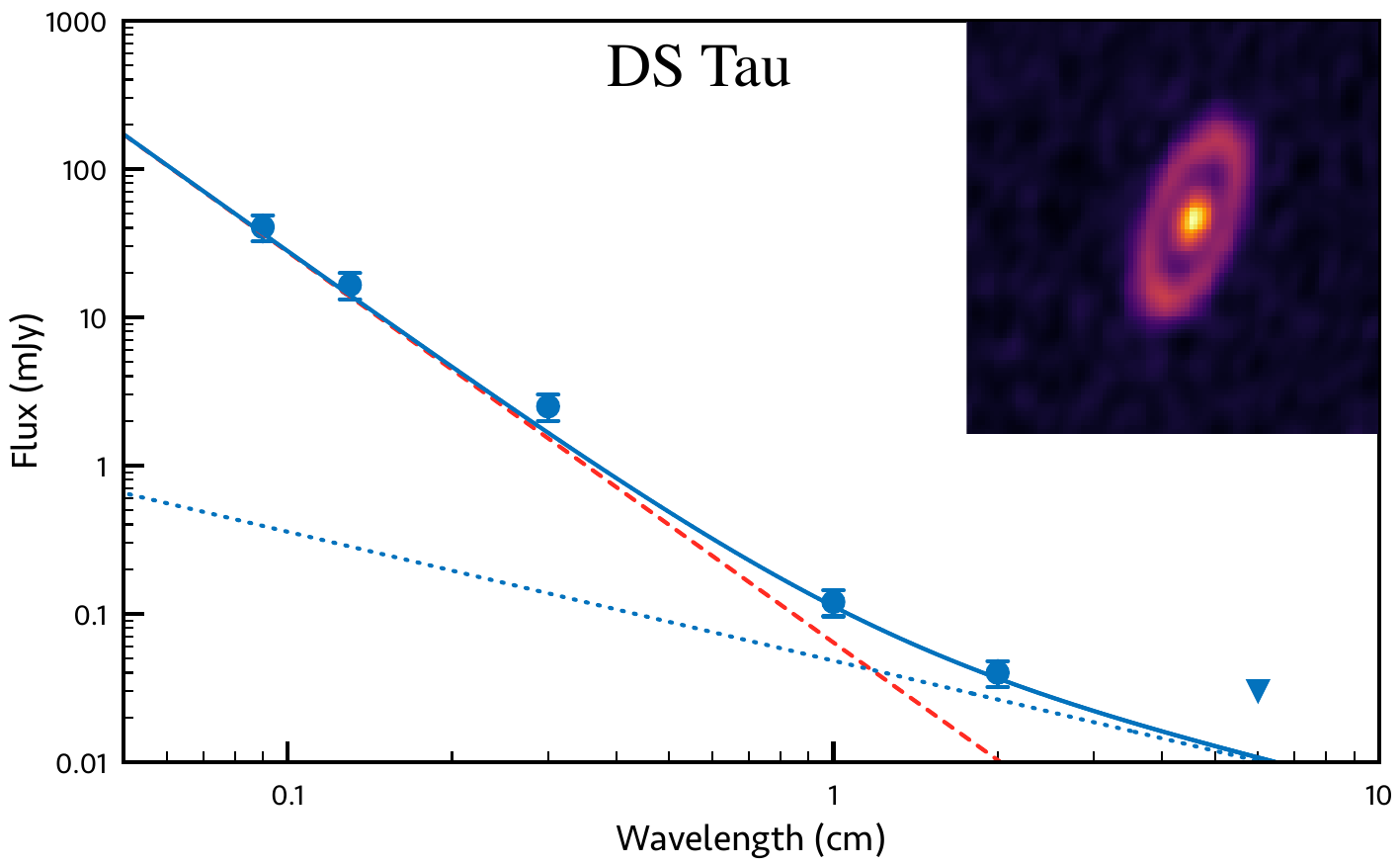}
   \includegraphics[width=5.5cm]{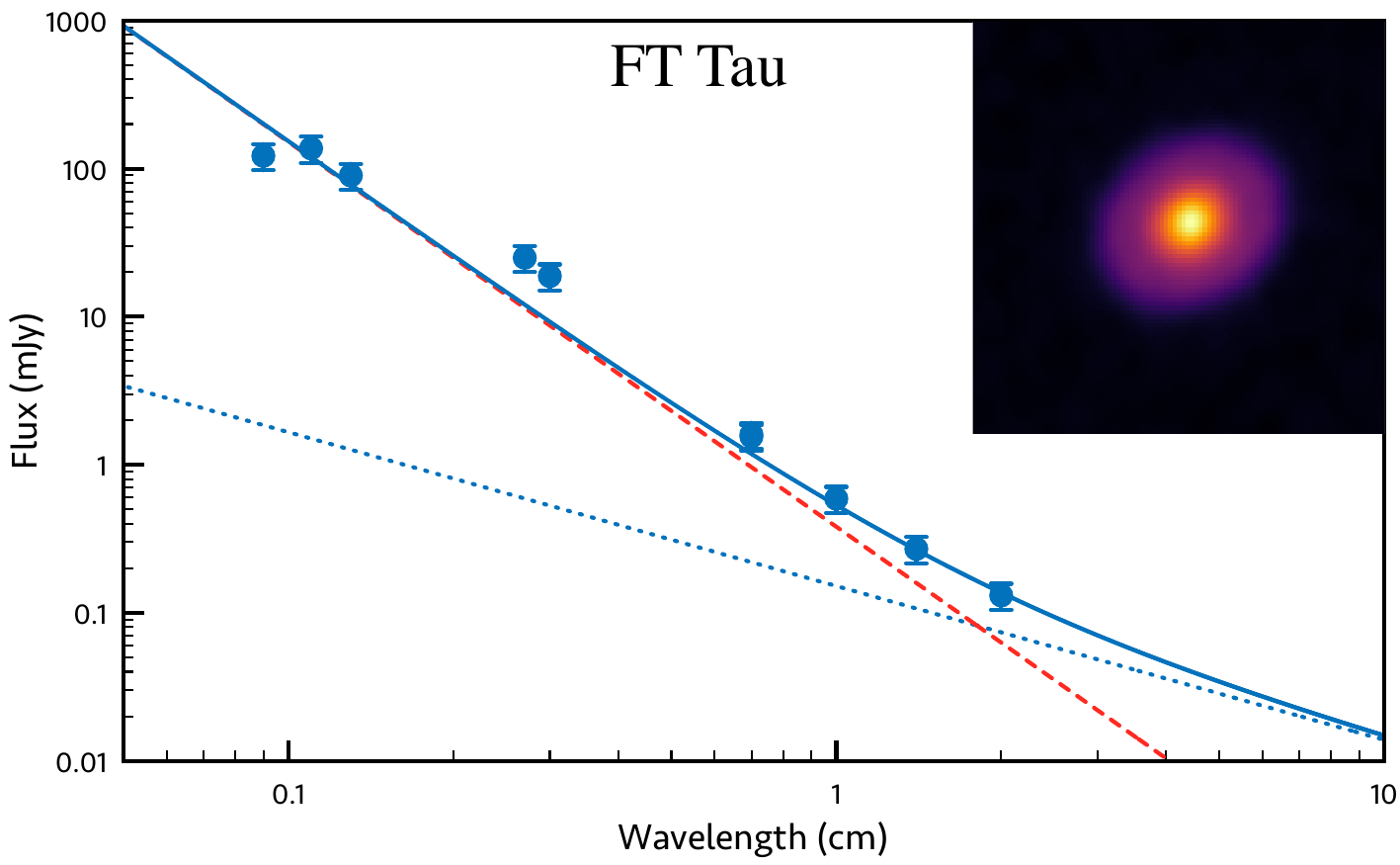}
   \includegraphics[width=5.5cm]{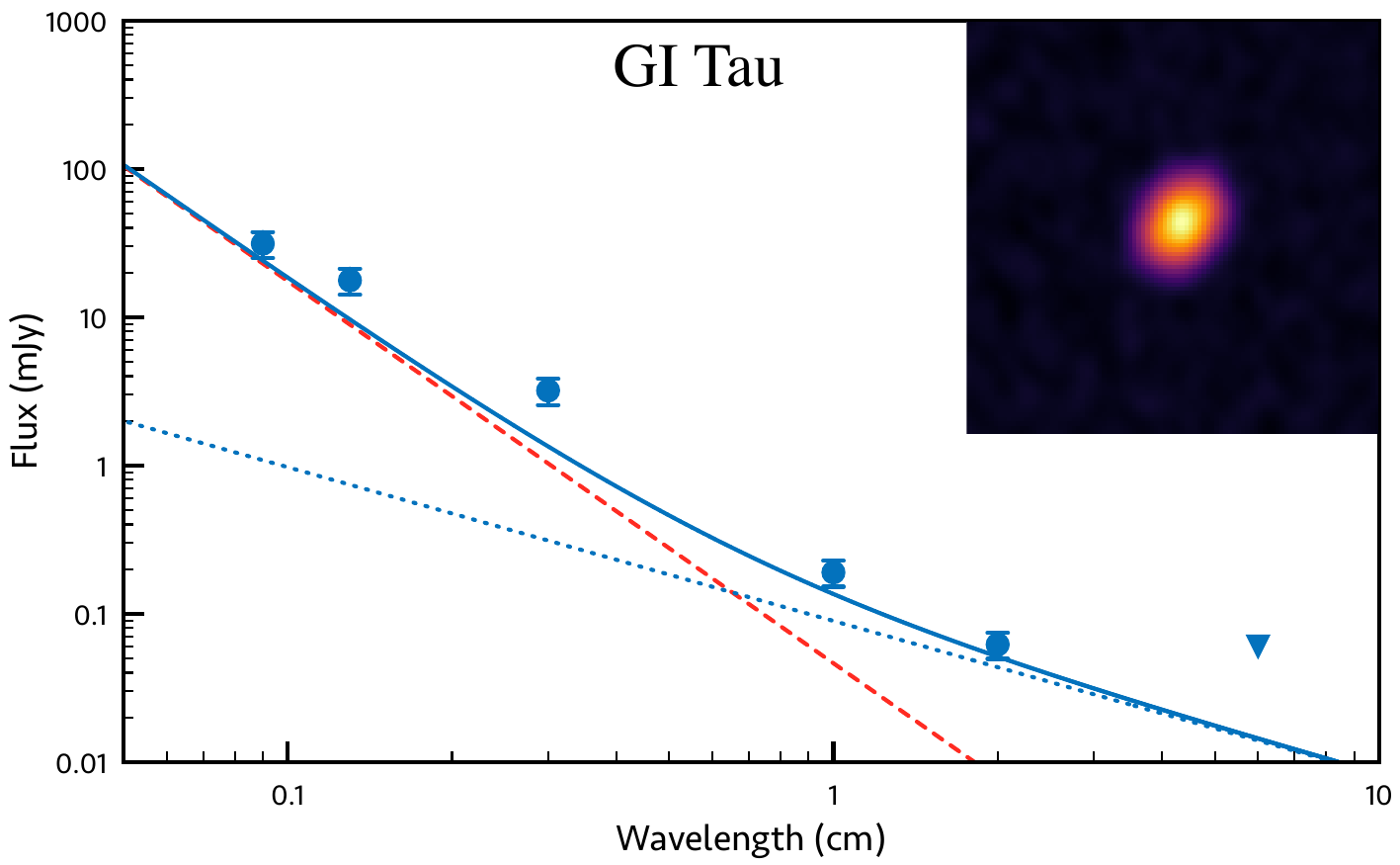}
   \includegraphics[width=5.5cm]{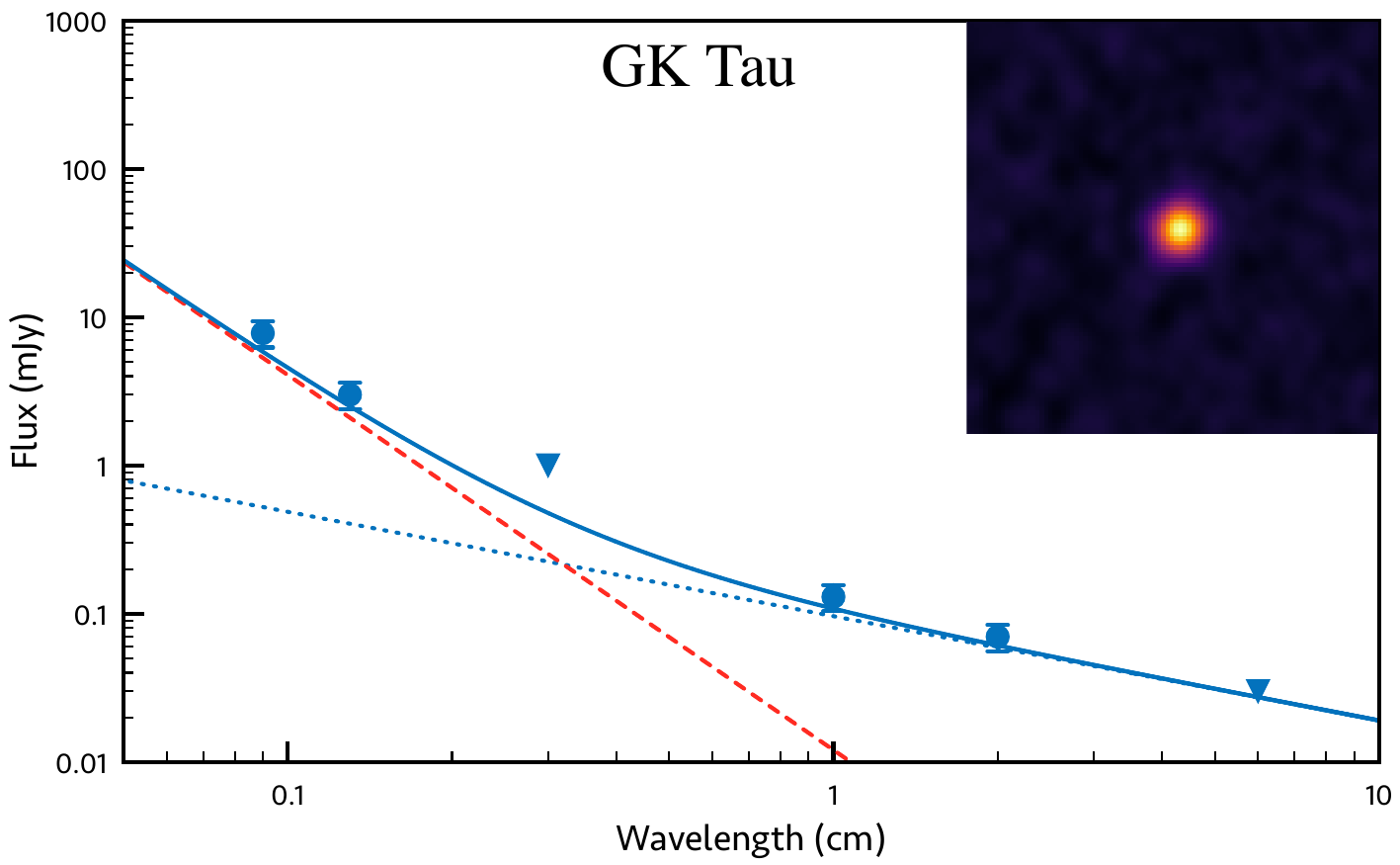}
   \includegraphics[width=5.5cm]{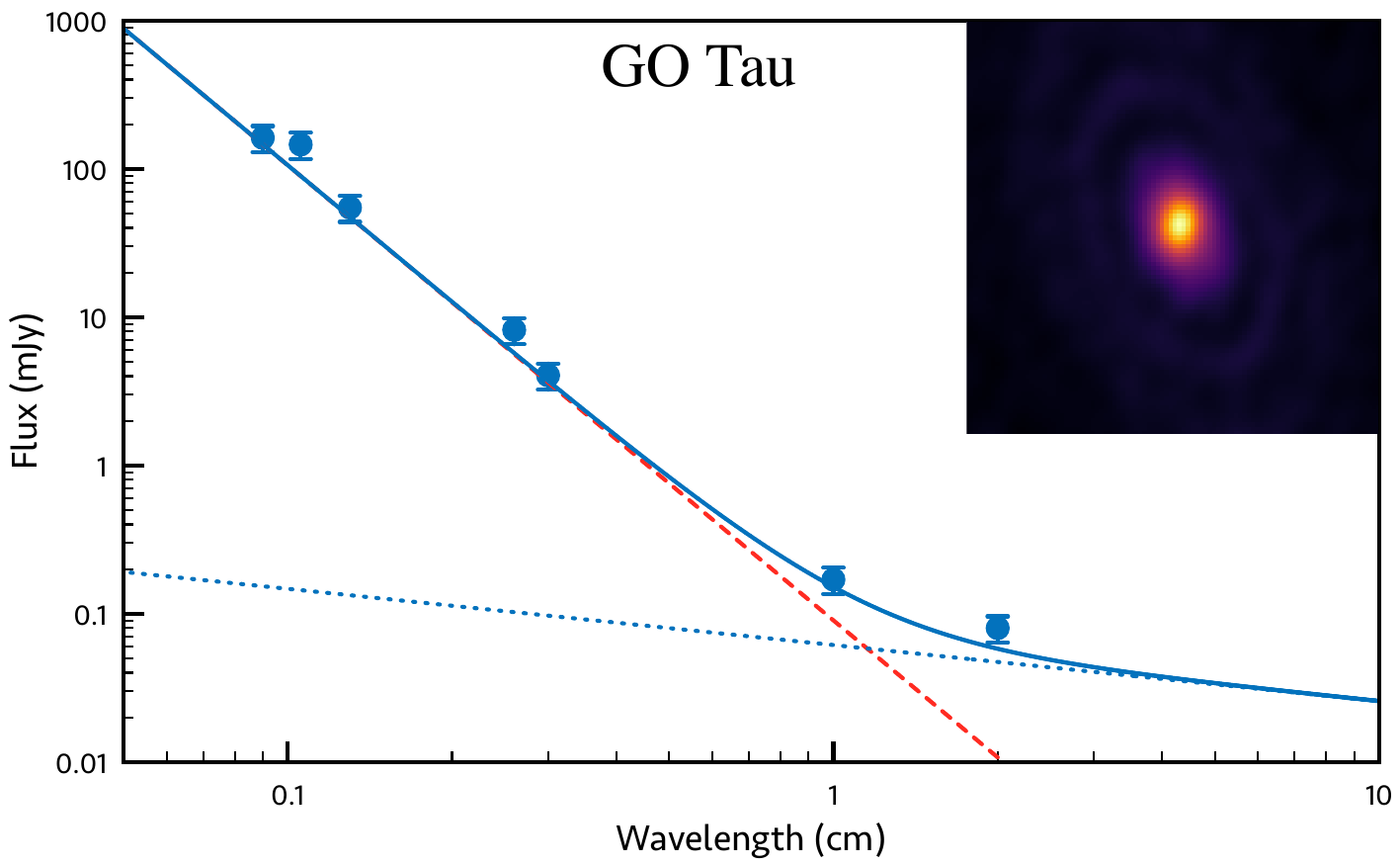}
   \includegraphics[width=5.5cm]{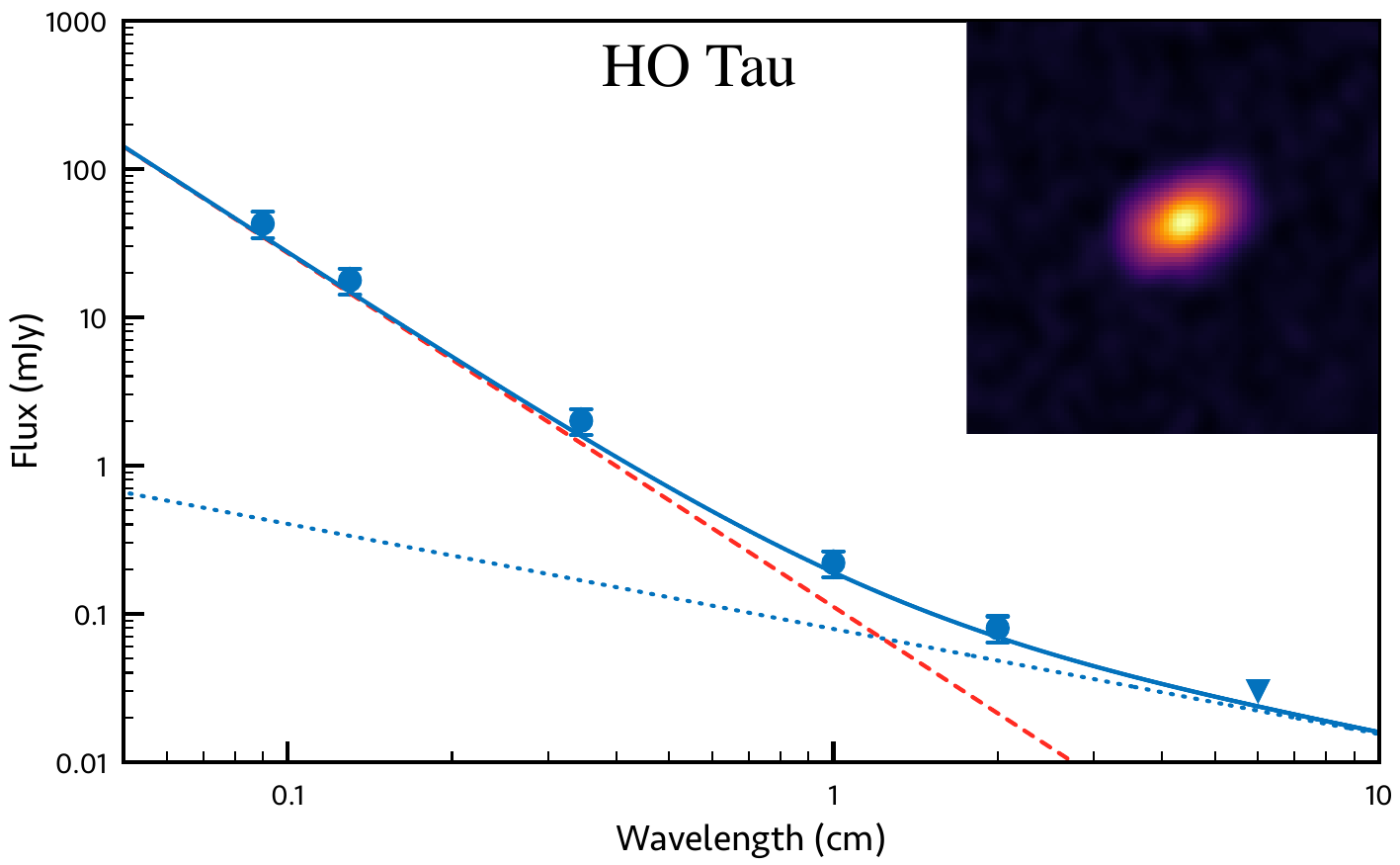}
   \includegraphics[width=5.5cm]{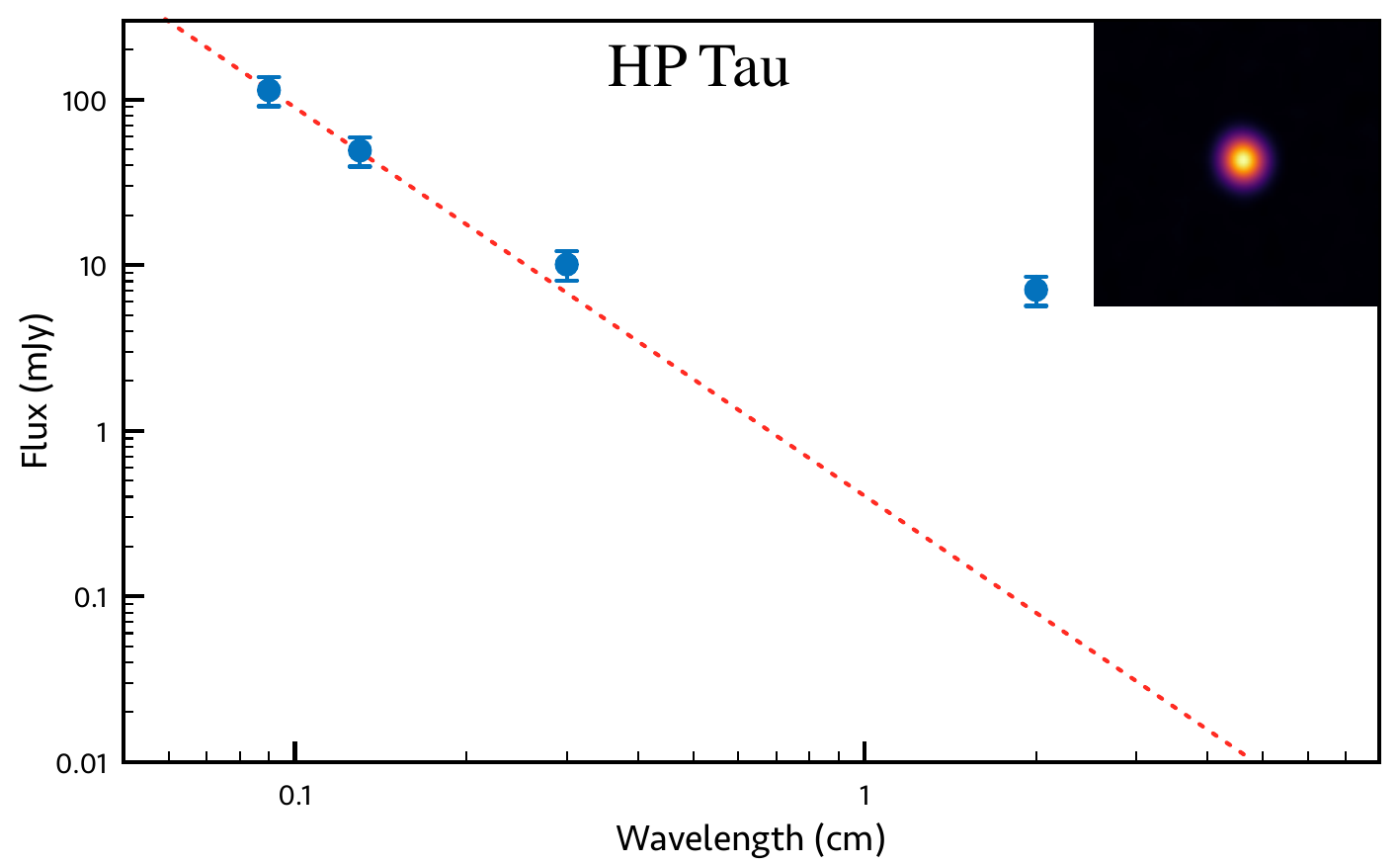}
   \includegraphics[width=5.5cm]{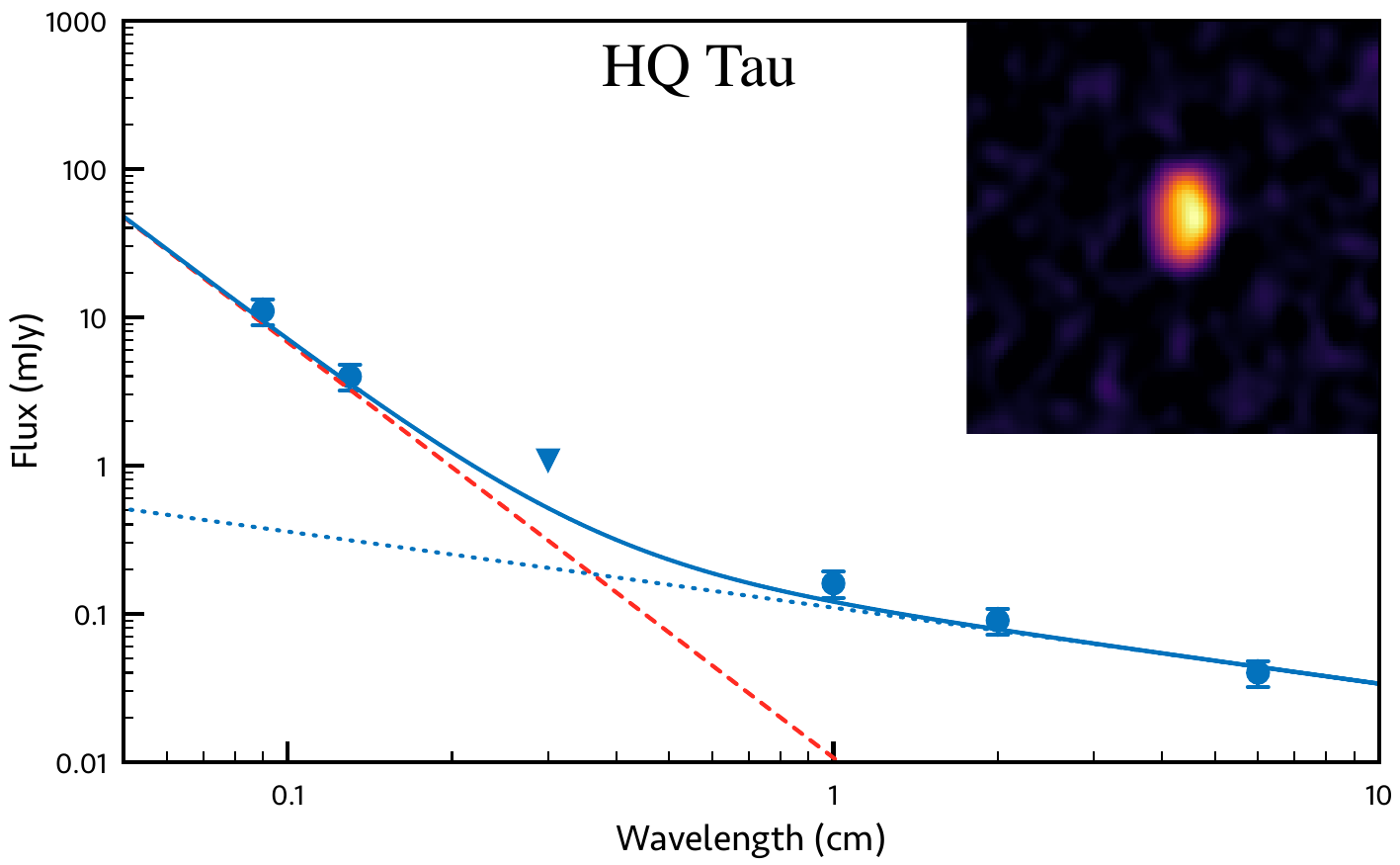}
   \includegraphics[width=5.5cm]{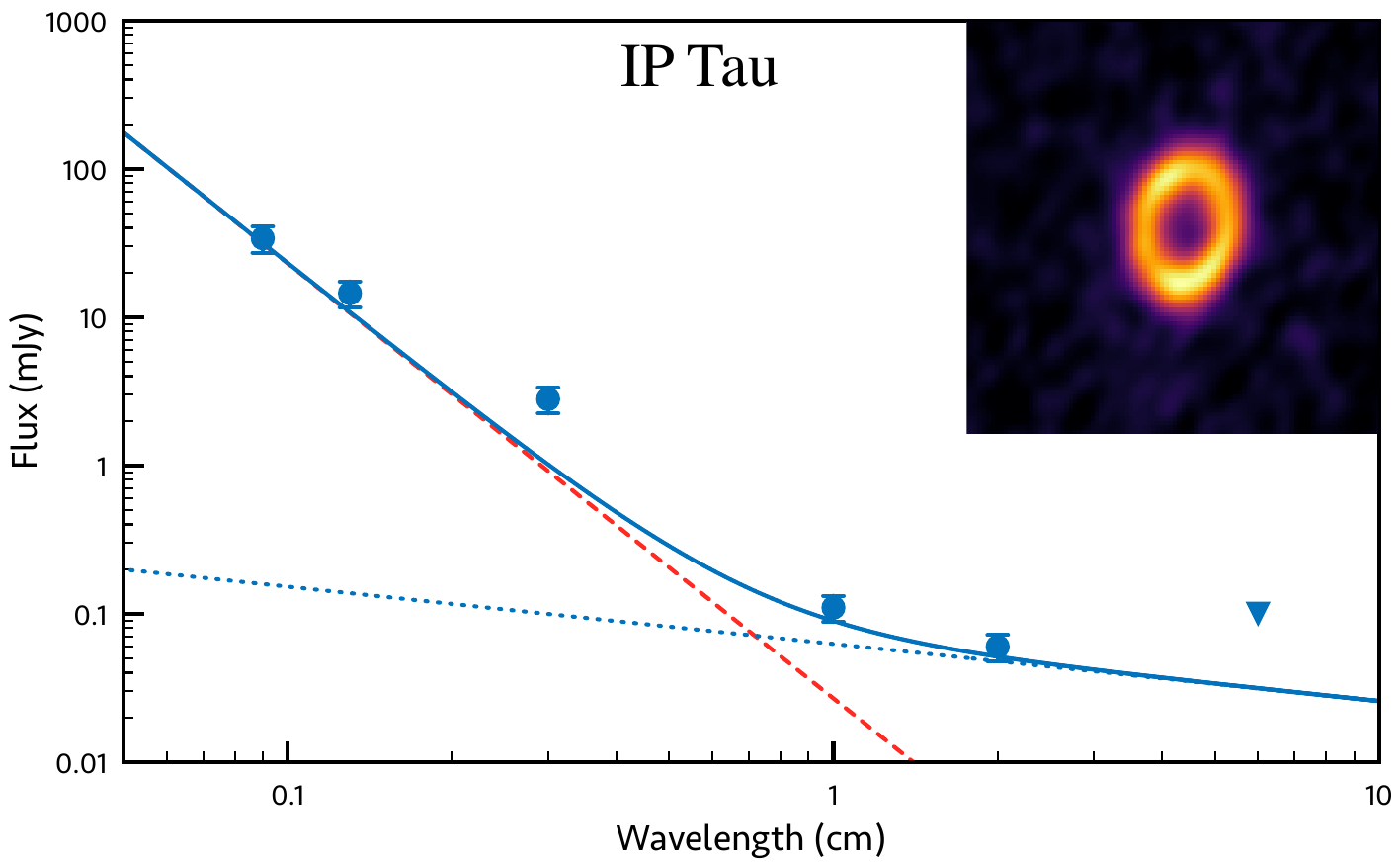}
   \includegraphics[width=5.5cm]{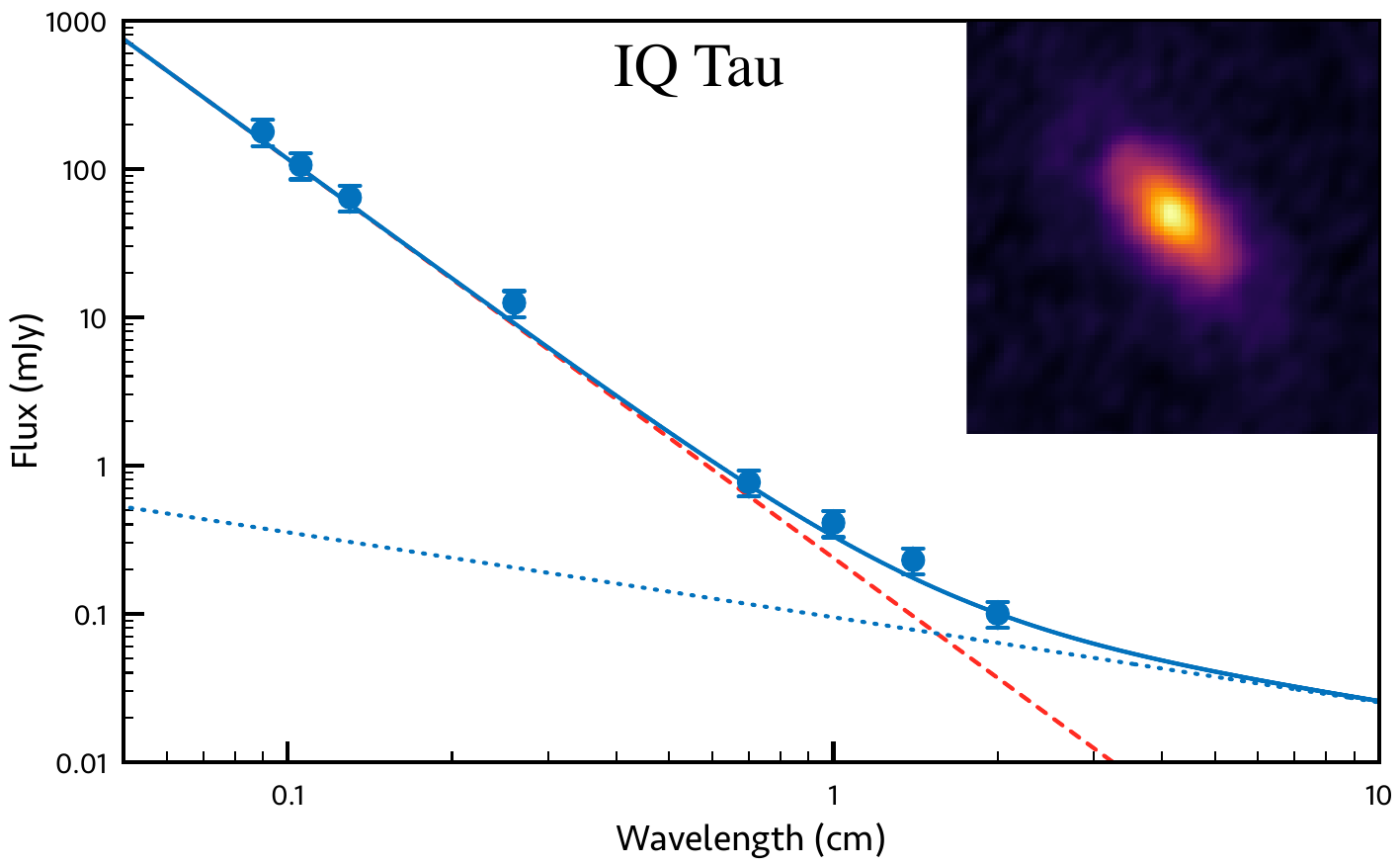}
   \includegraphics[width=5.5cm]{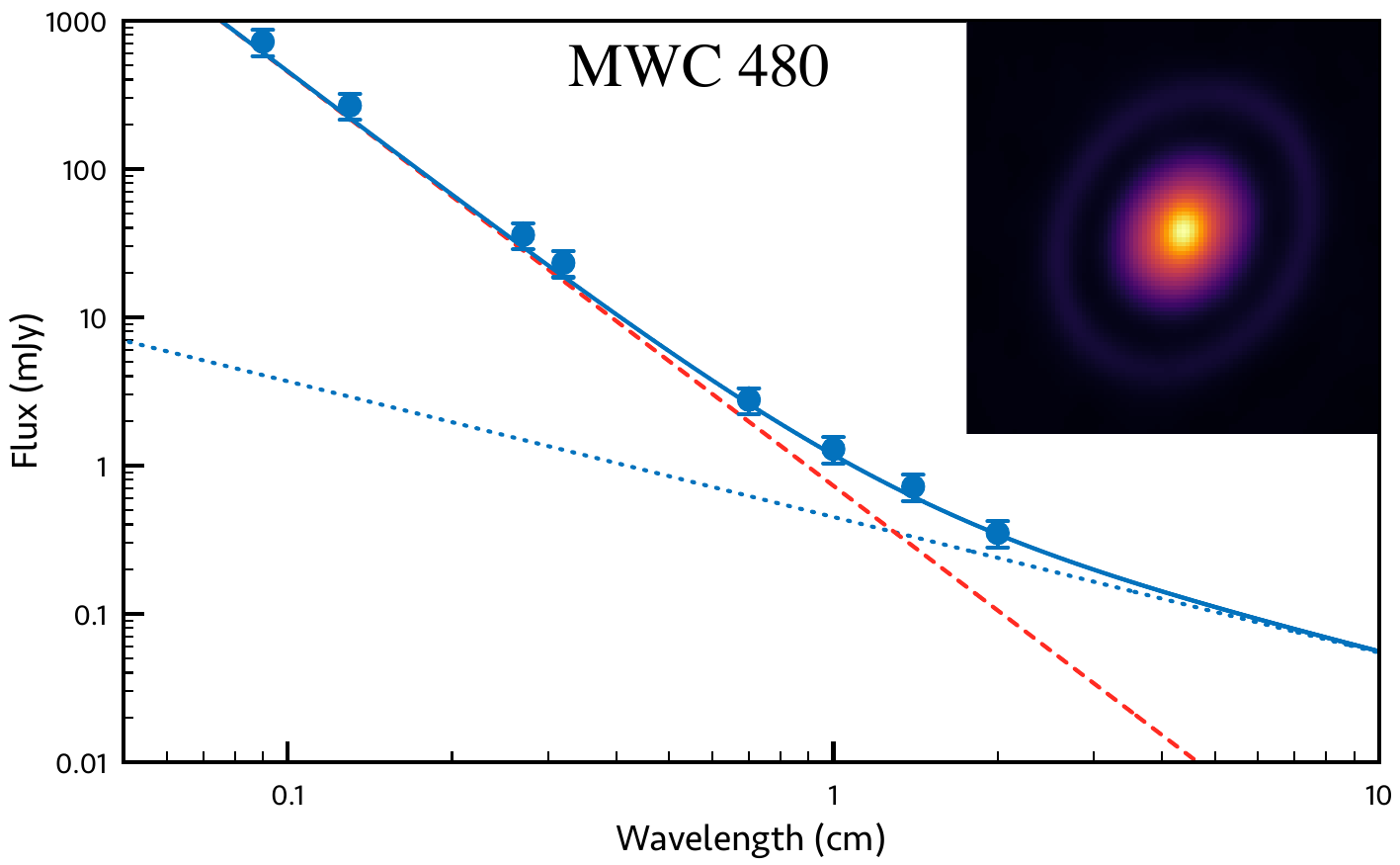}
   \includegraphics[width=5.5cm]{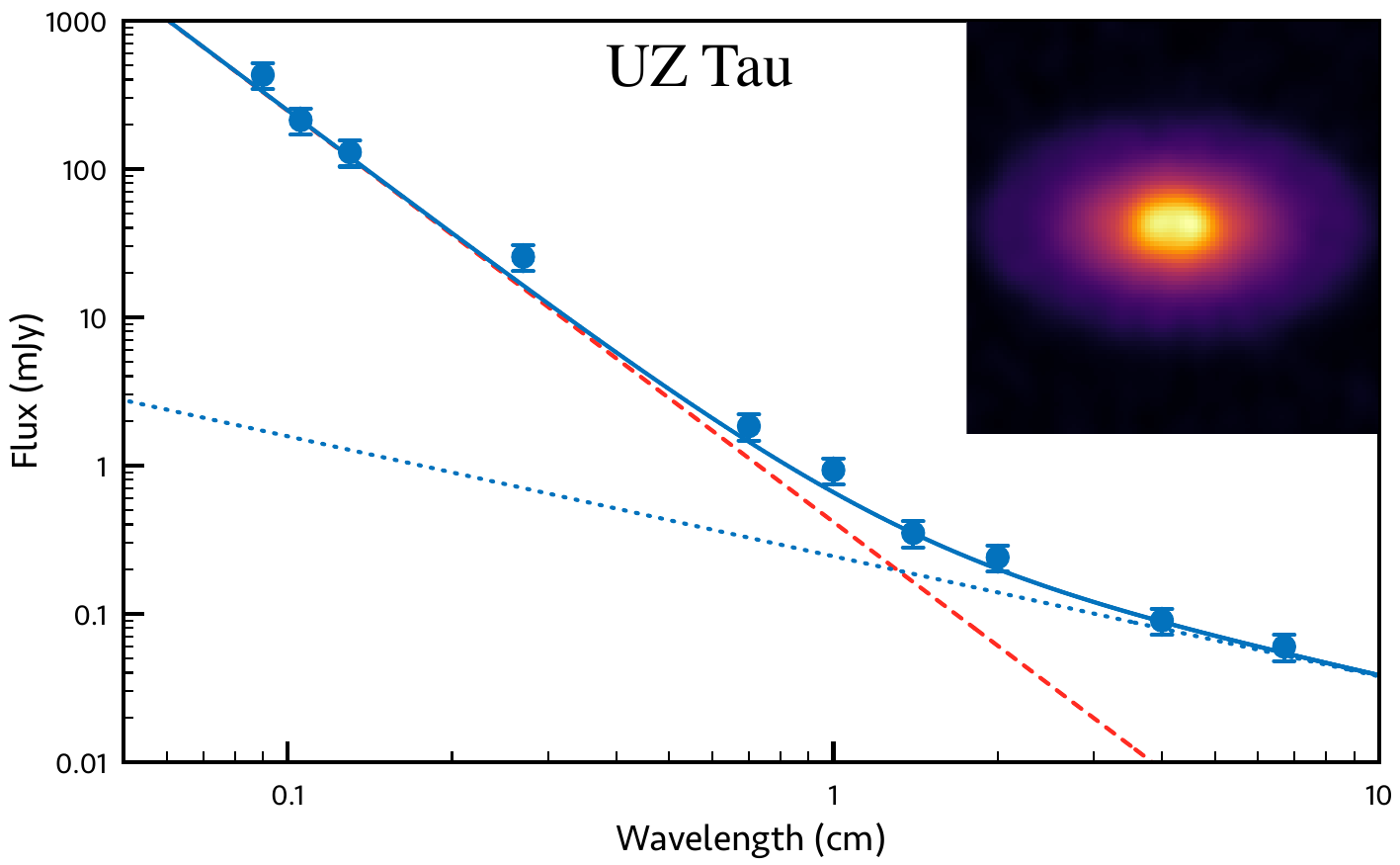}
   \includegraphics[width=5.5cm]{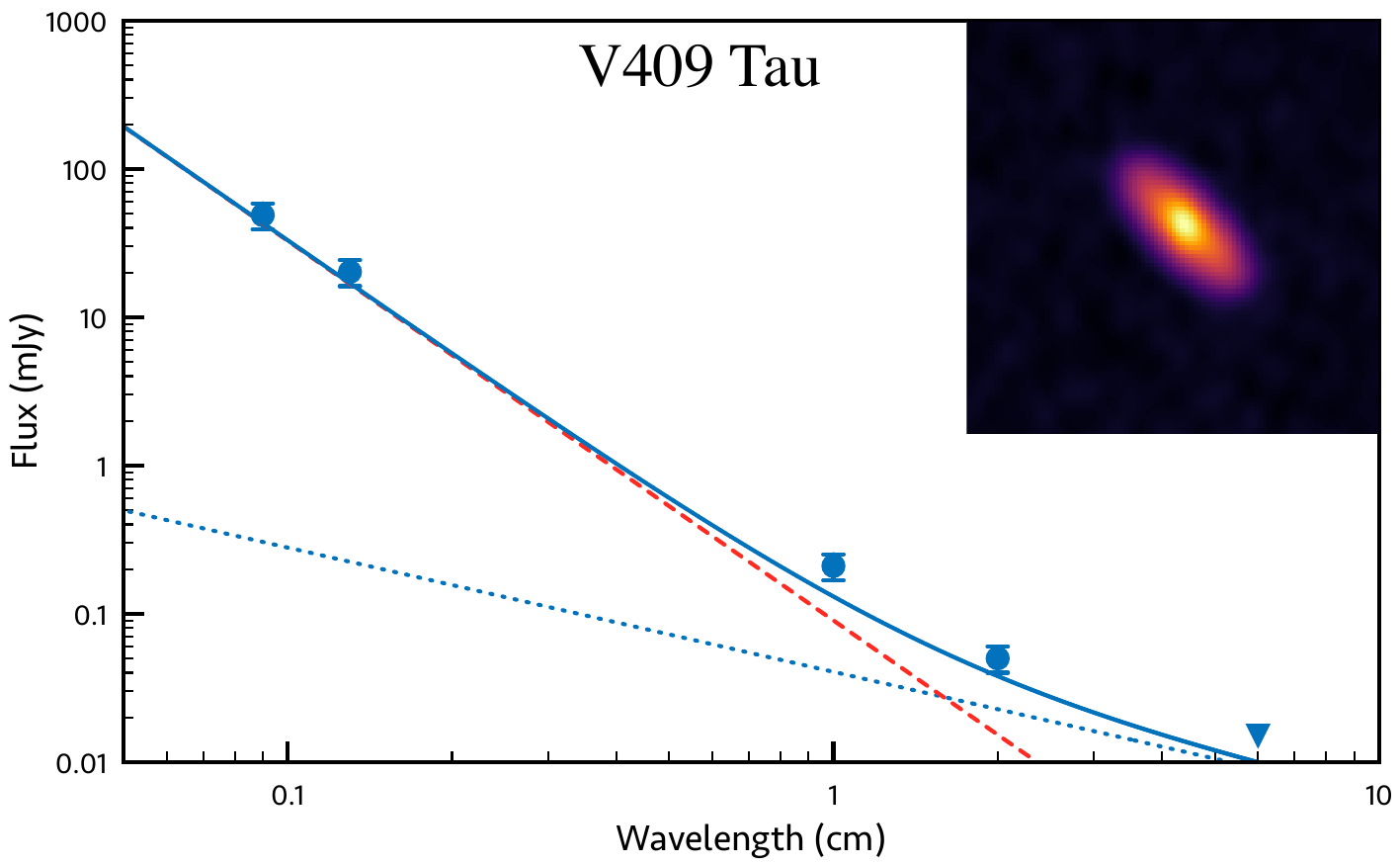}
   \includegraphics[width=5.5cm]{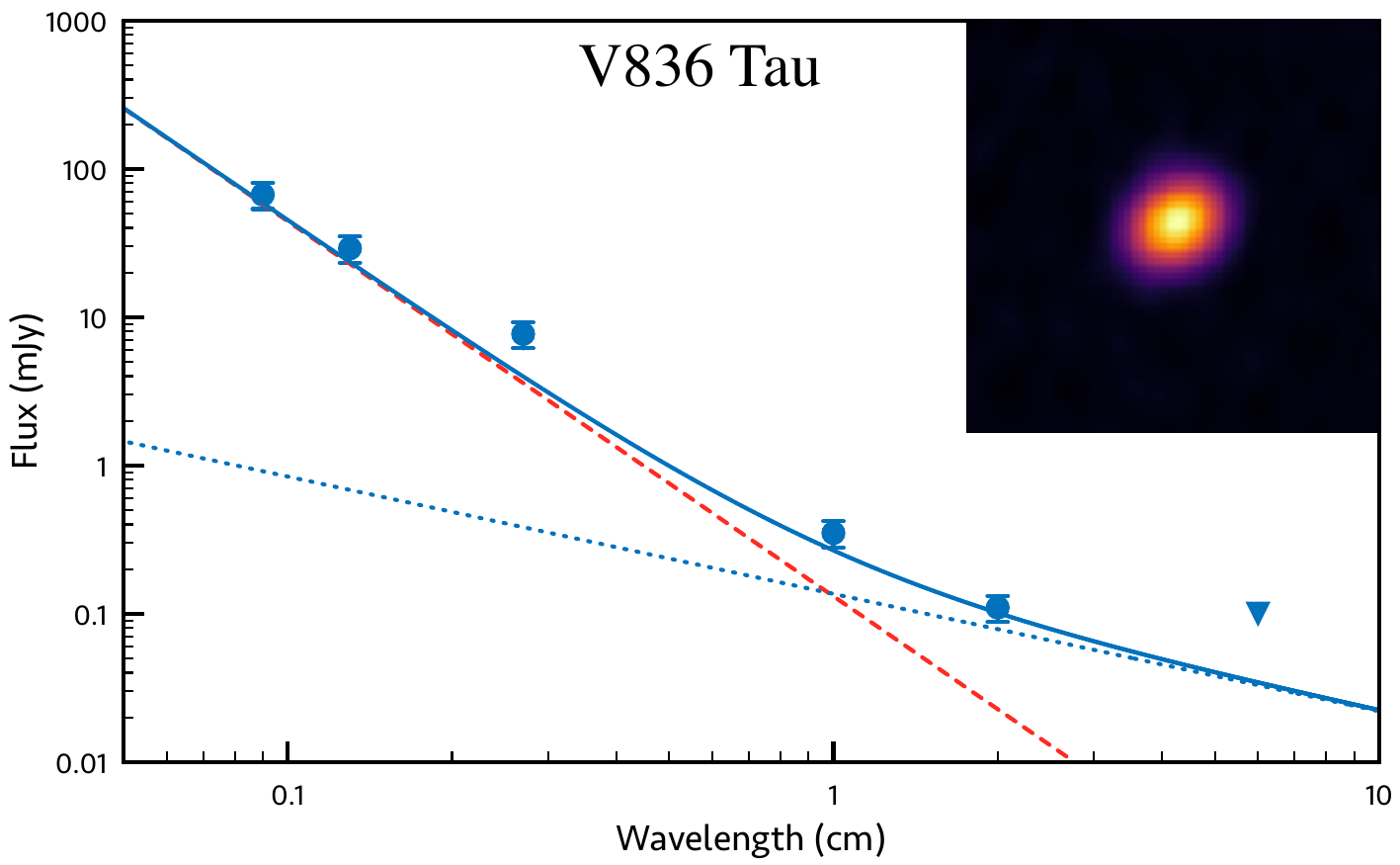}
   \caption{SED of the entire sample. The inset image is the ALMA continuum image by \citet{Long2019}. See Fig.\,\ref{fig:sed_illustrative} for details. The SED of HP Tau is not fitted because of the poor photometry available.}
              \label{fig:all_seds}%
    \end{figure*}

The VLA images described in Sect.\,\ref{sec:overview} are shown in Fig.\,\ref{fig:images_ku}. The SED fitting described in Sect.\,\ref{sec:sed} yields the parameters listed in Table \ref{table:sed_fit}. All SEDs and relative fits are shown in Fig.\,\ref{fig:all_seds}. Here we give a brief description of the individual cases referencing the ALMA images by \citet{Long2018, Long2019} and the SPHERE images of the scattered light by \citet{Garufi2024}.

BP Tau hosts one of the largest compact disks from ALMA with a peculiarly flat inner disk emission. The SPHERE image is clearly asymmetric, with one half of the disk significantly brighter than the other half, which is suggestive of large-angle shadowing. It is the only firm detection in the C band from our survey. Not surprisingly then, the spectral index of the free-free emission is the flattest of the sample ($\alpha_{\rm gas}=0.3$).

CIDA 9 has one of the two disks of the sample with a large cavity (with the other being IP Tau). It has a companion at 2.3\arcsec\ separation that in turns hosts a disk detected by ALMA \citep{Manara2019} but undetected in our Ku-band VLA image. The primary disk shows the second weakest free-free emission, after V409 Tau. It also has the highest dust spectral index between 1.3 mm and 2 cm (3.1) while the absence of any image at 1 cm prevents from constraining the other relevant indices.

DL Tau is surrounded by one of the three largest disks from the ALMA continuum emission that is nonetheless barely detected by SPHERE since self-shadowed. Multiple rings are visible in the ALMA image. It shows very high accretion and large free-free emission.

DN Tau hosts a standard disk within the sample. The relatively large disk seen by ALMA is barely detected in scattered light. The free-free emission and accretion rates are relatively low.

DO Tau is accompanied by one of the three relatively massive, compact disks (along with DQ Tau and DR Tau). It is a prototypical example of dust spectral index increasing with the wavelength (from $\alpha_{\rm 1mm}=2.1$ to $\alpha_{\rm 2cm}=3.2$). The source is enshrouded in a complex environment with evidence of extended outflows and accreting streamers \citep{Huang2022}. The large free-free emission (fourth in the sample), accretion rate (fourth), and near-IR excess (first) all indicate a connection between the environmental and circumstellar processes in act around this source.

DQ Tau is a spectroscopic binary with a stellar mass dynamically measured in 1.2 M$_\odot$ \citep{Czekala2016}. This estimate marginally differs from the photometric value reported in Table \ref{table:properties} (0.5+0.5 M$_\odot$). The disk is another good example of relatively massive though compact disk. The free-free emission constrained by our model is the highest of the sample (HP Tau is not treated by the model). However, in this case we choose to constrain the $\alpha_{\rm gas}$ of the model to 0.6 to prevent the model from determining a very steep value (1.6) that would result in an unnatural 35\% of the total emission at 1.33 mm being due to free-free emission. The final fit is much more uncertain than that of the other objects and that most likely reflects the complex environment of the binary system described in Sect.\,\ref{sec:discussion_cases}.  

DR Tau has the most massive of the compact disks. Similarly to DO Tau, several arc-like structures in likely interaction with the star are detected around the disk \citep{Mesa2022}. Another analogy with DO Tau is the presence of a high free-free emission, accretion rate, and near-IR excess (third, fifth, and second highest in the sample).

DS Tau hosts a disk with a central component surrounded by a single ring that contains approximately 80\% of the total flux. This disk is one of the three mildly inclined disks (60\degree--70\degree, along with IQ Tau and V409 Tau), and is barely detected by SPHERE.

\begin{table}
\caption{SED fit parameters.}             
\label{table:sed_fit}
\centering              
\begin{tabular}{l c c c c}  
\hline\hline              
Target & $S_{0}^{\rm dust}$ (mJy) & $\alpha_{\rm dust}$ & $S_{0}^{\rm gas}$ (mJy) & $\alpha_{\rm gas}$ \\ 
\hline 
   \smallskip
   BP Tau & 39.9$^{+3.8} _{-18.2}$ & 2.8$^{+0.7} _{-0.3}$ & 0.07$^{+0.02} _{-0.04}$ & 0.3$^{+1.1} _{-0.6}$ \\ 
   \smallskip 
   CIDA 9 & 24.1$^{+3.7} _{-10.9}$ & 2.8$^{+0.6} _{-0.5}$ & 0.02$^{+0.01} _{-0.02}$ & 0.6$^{+1.0} _{-0.9}$ \\ 
   \smallskip
   DL Tau & 155.5$^{+11.0} _{-29.2}$ & 2.9$^{+0.3} _{-0.2}$ & 0.10$^{+0.04} _{-0.07}$ & 0.8$^{+0.8} _{-1.3}$ \\  
   \smallskip
   DN Tau & 74.5$^{+8.0} _{-13.6}$ & 2.6$^{+0.3} _{-0.1}$ & 0.07$^{+0.04} _{-0.04}$ & 0.8$^{+0.9} _{-1.2}$ \\ 
   \smallskip
   DO Tau & 92.5$^{+13.4} _{-23.4}$ & 2.6$^{+0.5} _{-0.2}$ & 0.16$^{+0.05} _{-0.10}$ & 1.0$^{+0.8} _{-1.2}$ \\  
   \smallskip
   DQ Tau & 57.7$^{+18.7} _{-19.1}$ & 2.2$^{+0.9} _{-0.3}$ & 0.26$^{+0.11} _{-0.16}$ & 0.6 \\ 
   \smallskip
   DR Tau & 103.8$^{+9.9} _{-18.7}$ & 2.8$^{+0.4} _{-0.2}$ & 0.10$^{+0.03} _{-0.06}$ & 1.1$^{+0.7} _{-1.4}$ \\  
   \smallskip
   DS Tau & 13.8$^{+2.1} _{-5.4}$ & 2.6$^{+0.7} _{-0.4}$ & 0.03$^{+0.01} _{-0.03}$ & 0.9$^{+0.7} _{-1.2}$ \\ 
   \smallskip
   FT Tau & 76.7$^{+9.3} _{-34.5}$ & 2.6$^{+0.5} _{-0.2}$ & 0.07$^{+0.04} _{-0.04}$ & 1.0$^{+0.7} _{-1.3}$ \\  
   \smallskip
   GI Tau & 8.9$^{+2.6} _{-3.8}$ & 2.6$^{+0.8} _{-0.5}$ & 0.04$^{+0.02} _{-0.02}$ & 1.0$^{+0.6} _{-1.3}$ \\ 
   \smallskip
   GK Tau & 2.1$^{+0.7} _{-1.0}$ & 2.5$^{+0.9} _{-0.8}$ & 0.06$^{+0.01} _{-0.04}$ & 0.7$^{+0.5} _{-1.1}$ \\  
   \smallskip
   GO Tau & 47.2$^{+6.6} _{-13.5}$ & 3.0$^{+0.6} _{-0.4}$ & 0.05$^{+0.02} _{-0.03}$ & 0.4$^{+1.2} _{-0.8}$ \\ 
   \smallskip
   HO Tau & 14.4$^{+2.8} _{-6.0}$ & 2.4$^{+0.8} _{-0.3}$ & 0.05$^{+0.02} _{-0.03}$ & 0.7$^{+1.0} _{-1.1}$ \\  
   \smallskip
   HQ Tau & 3.2$^{+1.1} _{-1.2}$ & 2.8$^{+0.6} _{-0.9}$ & 0.08$^{+0.02} _{-0.03}$ & 0.5$^{+0.7} _{-0.9}$ \\  
   \smallskip
   IP Tau & 10.6$^{+3.9} _{-3.8}$ & 2.9$^{+0.6} _{-0.7}$ & 0.05$^{+0.01} _{-0.03}$ & 0.6$^{+0.9} _{-0.5}$ \\ 
   \smallskip
   IQ Tau & 57.5$^{+4.3} _{-14.1}$ & 2.7$^{+0.4} _{-0.2}$ & 0.06$^{+0.03} _{-0.03}$ & 0.6$^{+1.1} _{-1.0}$ \\
   \smallskip
   MWC 480 & 217.0$^{+38.4} _{-92.3}$ & 2.8$^{+0.5} _{-0.3}$ & 0.24$^{+0.08} _{-0.14}$ & 0.9$^{+0.8} _{-1.2}$ \\  
   \smallskip
   UZ Tau E & 119.0$^{+7.6} _{-23.9}$ & 2.8$^{+0.4} _{-0.2}$ & 0.14$^{+0.07} _{-0.05}$ & 0.8$^{+0.7} _{-0.4}$ \\ 
   \smallskip
   V409 Tau & 16.7$^{+2.7} _{-6.7}$ & 2.6$^{+0.7} _{-0.3}$ & 0.02$^{+0.02} _{-0.01}$ & 0.8$^{+0.9} _{-1.2}$ \\  
   \smallskip
   V836 Tau & 22.8$^{+4.4} _{-9.7}$ & 2.5$^{+0.8} _{-0.5}$ & 0.08$^{+0.02} _{-0.05}$ & 0.8$^{+0.9} _{-1.1}$ \\ 
\hline 
\end{tabular}
\tablefoot{Columns are target name, dust flux at 1.3 mm, dust spectral index, {ionized gas flux (free-free or synchrotron emission)} at 2 cm, and ionized gas spectral index. Quoted errors for each parameter are given by percentiles at 16 and 84, which correspond to 1-sigma errors for Gaussian probability distributions. The $\alpha_{\rm gas}$ of DQ Tau is assumed (see text).}
\end{table}

FT Tau is surrounded by a disk with a central component surrounded by a bright ring that accounts for about 60\% of the total flux (similarly to that of DS Tau).

GI Tau hosts a compact disk with low free-free emission. The disk is one of the least massive disks ever detected in scattered light.

GK Tau has the least massive disk of the sample, together with HQ Tau. It is undetected in scattered light. 

GO Tau is surrounded by a disk with a bright, central component and multiple rings further out that however only account for about 20\% of the total flux. In fact, the disk is not particularly massive (being averaged in the sample) given its extent (being the largest in the sample). The accretion rate is among the lowest of the sample. 

HO Tau has a low-mass compact disk with low free-free emission and accretion rate. The disk has no peculiar feature to be mentioned.

HP Tau hosts an ordinary, compact disk in the ALMA continuum. However, the source is surrounded by a very complex field of arc-like structure that are visible in the SPHERE image and may be due to a recent encounter with nearby companions \citep[see][]{Garufi2024}. The poor available photometry and the anomalously high 2-cm flux prevent any SED fitting.

HQ Tau is surrounded by the least massive disk of the sample, together with GK Tau. The disk is marginally detected in scattered light. 

IP Tau hosts one of the two disks of the sample with a resolved cavity (with the other being CIDA 9). The disk is relatively bright in scattered light given its very low dust mass, and the SPHERE image reveals evidence of asymmetric structures like spirals in correspondence of the ALMA ring. The source shows very low free-free emission and accretion rate. 

IQ Tau has a relatively inclined disk with multiple rings seen by ALMA that is barely visible in the SPHERE image since self-shadowed. Similarly to DL Tau and GO Tau, the ALMA rings account for a small fraction of the total flux.

MWC 480 is the most massive star of the sample (2.1 M$_\odot$) and hosts the most massive disk. The central disk component seen by ALMA is marginally detected by SPHERE as well. It shows the second strongest free-free emission (excluding HP Tau) and the largest mass accretion rate of the sample.

UZ Tau E is a binary star with a total dynamical mass of 1.3 M$_\odot$ \citep{Simon2000}, which is a very discrepant value from the photometric value of Table \ref{table:properties} (0.7 M$_\odot$, although both estimates are quite uncertain). The circumbinary disk exhibits a small, shallow cavity in ALMA encompassed by several rings. The disk is also clearly resolved by SPHERE. The companion UZ Tau W is by itself a binary with 0.4\arcsec\ separation lying at a separation of 3.5\arcsec\ from the primary. Both components of UZ Tau W host disks with a similar inclination to that of UZ Tau E \citep{Manara2019} that are resolved by ALMA and SPHERE, and are detected but unresolved in our VLA Ku-band image.  

V409 Tau is surrounded by a compact, inclined disk that is also clearly imaged by SPHERE. It shows the lowest free-free emission, accretion rate, and near-IR excess of the sample.

V836 Tau hosts a compact disk with no notable peculiarity. Yet, the free-free emission is strong (fifth highest value in the sample).

\begin{figure}
    \centering
    \includegraphics[width=8.5cm]{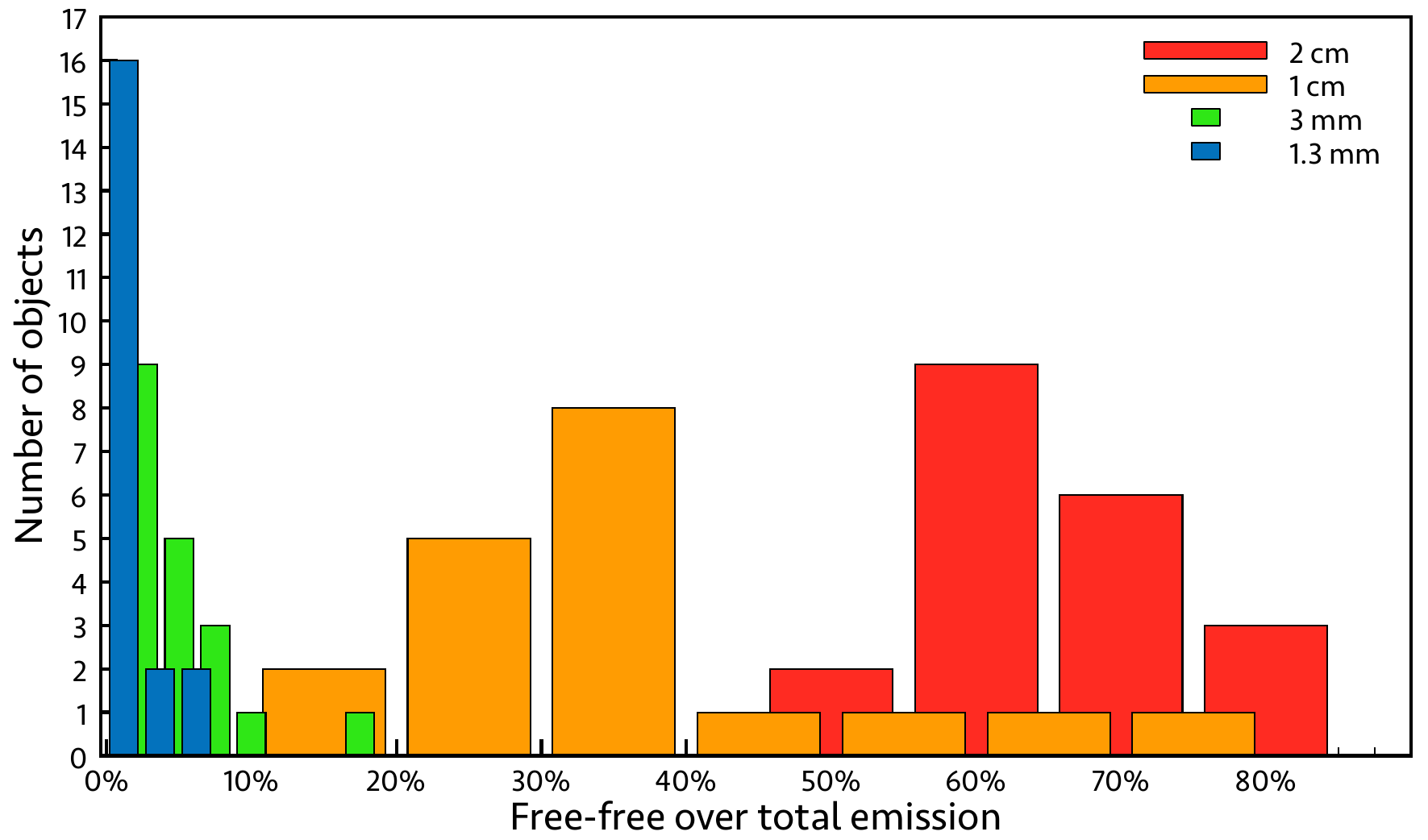}
    \caption{Free-free over total emission. The fraction of free-free emission over total emission that is obtained from our SED fitting is shown at 1.3 mm, 3 mm, 1 cm, and 2 cm.}
    \label{fig:ff_fraction}
\end{figure}

 \begin{table*}
 \caption{Setup and properties of the VLA observations.}             
 \label{table:setup}
 \centering
 \begin{tabular}{l c c c c c c c} 
 \hline              
 Target & Waveband & Project & Observation & Beam size & Beam  & RMS & Measured \\
  & (GHz) & & date &  (\arcsec) & angle ($\degree$) & (mJy/beam) & flux (mJy) \\
 \hline                     
 BP Tau & Ka (33) & 20A-373 & 8-10-20 & 0.18$\times$0.17 & -53 & 0.007 & 0.26 $\pm$ 0.02 \\  
 BP Tau & Ku (15) & 20A-373 & 2-3-20 & 1.40$\times$1.32 & -21 & 0.004 & 0.10 $\pm$ 0.01 \\  
   BP Tau & C (6) & 20A-373 & 20-2-20 & 3.75$\times$2.97 & -16 & 0.007 & 0.07 $\pm$ 0.02 \\  
   CIDA 9 & Ku (15) & 20A-373 & 8-3-20 & 1.78$\times$1.37 & -74 & 0.004 & 0.03 $\pm$ 0.01 \\
   CIDA 9 & C (6) & 20A-373 & 25-2-20 & 3.23$\times$2.91 & 2 & 0.006 & <0.02 \\
   DL Tau & Q (43) & 21B-267 & 14-10-21 & 0.18$\times$0.15 & 65 & 0.013 & 1.81 $\pm$ 0.10 \\  
   DL Tau & Ka (33) & 21B-267 & 20-11-21 & 0.24$\times$0.19  & -56 & 0.015 & 0.67 $\pm$ 0.07 \\  
   DL Tau & K (22) & 21B-267 & 13-12-21 & 0.43$\times$0.31 & -51 & 0.008 & 0.29 $\pm$ 0.02 \\  
   DL Tau & Ku (15) & 20A-373 & 17-2-20 & 1.39$\times$1.20 & -36  & 0.005 & 0.17 $\pm$ 0.02 \\  
   DN Tau & Q (43) & 21B-267 & 1-11-21 & 0.16$\times$0.15 & -14 & 0.012 & 1.18 $\pm$ 0.06 \\ 
   DN Tau & Ka (33) & 21B-267 & 11-12-21 & 0.25$\times$0.20 & 89 & 0.015 & 0.70 $\pm$ 0.02 \\ 
   DN Tau & K (22) & 21B-267 & 30-12-21 & 0.33$\times$0.28 & -8 & 0.009 & 0.24 $\pm$ 0.01 \\ 
   DN Tau & Ku (15) & 20A-373 & 19-2-20 & 1.44$\times$1.24 & -7 & 0.005 & 0.14 $\pm$ 0.02 \\ 
   DO Tau & Q (43) & 21B-267 & 18-12-21 & 0.22$\times$0.16 & -53 & 0.014 & 2.18 $\pm$ 0.04 \\ 
   DO Tau & Ka (33) & 21B-267 & 18-12-21 & 0.21$\times$0.18 & -7 & 0.017 & 0.95 $\pm$ 0.07 \\ 
   DO Tau & K (22) & 21B-267 & 31-12-21 & 0.47$\times$0.31 & -55 & 0.009 & 0.41 $\pm$ 0.02 \\ 
   DO Tau & Ku (15) & 20A-373 & 15-2-20 & 1.49$\times$1.25 & -30 & 0.005 & 0.23 $\pm$ 0.02 \\ 
   DQ Tau & Q (43) & 21B-267 & 18-12-21 & 0.17$\times$0.16 & 51 & 0.014 & 2.43 $\pm$ 0.09 \\ 
   DQ Tau & Ka (33) & 21B-267 & 12-12-21 & 0.31$\times$0.20 & -45 & 0.015 & 1.54 $\pm$ 0.07 \\ 
   DQ Tau & K (22) & 21B-267 & 21-12-21 & 0.33$\times$0.27 & -3 & 0.012 & 31.24 $\pm$ 0.05 \\ 
   DQ Tau & Ku (15) & 20A-373 & 15-2-20 & 1.80$\times$1.26 & -42 & 0.005 & 0.41 $\pm$ 0.02 \\ 
   DR Tau & Q (43) & 21B-267 & 18-12-21 & 0.17$\times$0.15 & 33 & 0.014 & 1.31 $\pm$ 0.09 \\ 
   DR Tau & Ka (33) & 21B-267 & 19-12-21 & 0.30$\times$0.20 & -45 & 0.016 & 0.45 $\pm$ 0.05 \\ 
   DR Tau & K (22) & 21B-267 & 21-12-21 & 0.33$\times$0.27 & -3 & 0.009 & 0.31 $\pm$ 0.03 \\ 
   DR Tau & Ku (15) & 20A-373 & 4-3-20 & 1.43$\times$1.26 & 7 & 0.004 & 0.15 $\pm$ 0.01 \\ 
   DS Tau & Ka (33) & 20A-373 & 17-10-20 & 0.27$\times$0.16 & 73 & 0.006 & 0.12 $\pm$ 0.02 \\ 
   DS Tau & Ku (15) & 20A-373 & 8-3-20 & 1.38$\times$1.33 & 27 & 0.004 & 0.04 $\pm$ 0.01  \\ 
   DS Tau & C (6) & 20A-373 & 25-2-20 & 3.18$\times$2.98 & -1 & 0.010 & <0.03 \\ 
   FT Tau & Q (43) & 21B-267 & 18-12-21 & 0.16$\times$0.16 & -53 & 0.011 & 1.55 $\pm$ 0.05 \\ 
   FT Tau & Ka (33) & 21B-267 & 11-12-21 & 0.25$\times$0.20 & 87 & 0.015 & 0.59 $\pm$ 0.05 \\ 
   FT Tau & K (22) & 21B-267 & 21-12-21 & 0.33$\times$0.28 & -10 & 0.010 & 0.27 $\pm$ 0.03 \\ 
   FT Tau & Ku (15) & 20A-373 & 29-2-20 & 1.53$\times$1.39 & 75 & 0.004 & 0.13 $\pm$ 0.01 \\ 
   GI Tau & Ka (33) & 20A-373 & 24-8-20 & 0.30$\times$0.17 & -68 & 0.009 & 0.19 $\pm$ 0.03 \\ 
   GI Tau & Ku (15) & 20A-373 & 2-3-20 & 1.57$\times$1.31 & -35 & 0.006 & 0.06 $\pm$ 0.02 \\ 
   GI Tau & C (6) & 20A-373 & 20-2-20 & 3.67$\times$2.78 & -5 & 0.028 & <0.06 \\ 
   GK Tau & Ka (33) & 20A-373 & 14-9-20 & 0.24$\times$0.17 & 59 & 0.007 & 0.13 $\pm$ 0.02 \\ 
   GK Tau & Ku (15) & 20A-373 & 17-2-20 & 1.50$\times$1.25 & -30 & 0.005 & 0.07 $\pm$ 0.02 \\ 
   GK Tau & C (6) & 20A-373 & 20-2-20 & 3.69$\times$2.76 & -4 & 0.008 & <0.03 \\
   GO Tau & Ka (33) & 20A-373 & 7-11-20 & 0.27$\times$0.08 & -66 & 0.010 & 0.17 $\pm$ 0.03 \\
   GO Tau & Ku (15) & 20A-373 & 15-2-20 & 1.64$\times$1.24 & -40 & 0.005 & 0.08 $\pm$ 0.02 \\
   HO Tau & Ka (33) & 20A-373 & 1-10-20 & 0.27$\times$0.17 & -64 & 0.008 & 0.22 $\pm$ 0.02 \\ 
   HO Tau & Ku (15) & 20A-373 & 19-2-20 & 1.46$\times$1.31 & -3 & 0.005 & 0.08 $\pm$ 0.02 \\ 
   HO Tau & C (6) & 20A-373 & 25-2-20 & 3.35$\times$3.14 & 17 & 0.006 & <0.03 \\ 
   HP Tau & Ku (15) & 20A-373 & 15-2-20 & 1.47$\times$1.24 & -30 & 0.005 & 7.09 $\pm$ 0.02 \\ 
   HQ Tau & Ka (33) & 20A-373 & 16-10-20 & 0.23$\times$0.17 & 70 & 0.007 & 0.16 $\pm$ 0.02 \\ 
   HQ Tau & Ku (15) & 20A-373 & 19-2-20 & 1.44$\times$1.21 & -3 & 0.005 & 0.09 $\pm$ 0.02 \\ 
   HQ Tau & C (6) & 20A-373 & 25-2-20 & 3.30$\times$3.15 & 17 & 0.006 & 0.04 $\pm$ 0.02 \\ 
   IP Tau & Ka (33) & 20A-373 & 2-10-20  & 0.22$\times$0.16 & 87 & 0.008 & 0.11 $\pm$ 0.02 \\ 
   IP Tau & Ku (15) & 20A-373 & 29-2-20 & 1.41$\times$1.37 & 87 & 0.004 & 0.06 $\pm$ 0.01 \\ 
   IP Tau & C (6) & 20A-373 & 20-2-20 & 3.69$\times$2.88 & -9 & 0.024 & <0.08 \\
   IQ Tau & Q (43) & 21B-267 & 13-12-21  & 0.20$\times$0.16 & -88 & 0.012 & 0.77 $\pm$ 0.07 \\  
   IQ Tau & Ka (33) & 21B-267 & 18-12-21 & 0.21$\times$0.18 & -6 & 0.015 & 0.41 $\pm$ 0.04 \\  
   IQ Tau & K (22) & 21B-267 & 21-12-21 & 0.34$\times$0.32 & 71 & 0.011 & 0.23 $\pm$ 0.05 \\  
   IQ Tau & Ku (15) & 20A-373 & 29-2-20 & 1.40$\times$1.31 & -14 & 0.004 & 0.10 $\pm$ 0.01 \\ 
   MWC 480 & Q (43) & 21B-267 & 12-12-21 & 0.18$\times$0.17 & 89 & 0.012 & 2.77 $\pm$ 0.07 \\  
   MWC 480 & Ka (33) & 21B-267 & 10-11-21 & 0.22$\times$0.19 & -55 & 0.017 & 1.29 $\pm$ 0.07 \\  
   MWC 480 & K (22) & 21B-267 & 21-12-21 & 0.38$\times$0.30 & -83 & 0.010 & 0.72 $\pm$ 0.03 \\  
   MWC 480 & Ku (15) & 20A-373 & 8-3-20 & 1.44$\times$1.36 & 86 & 0.004 & 0.36 $\pm$ 0.01 \\
   \hline 
   \end{tabular}
   \end{table*}

\begin{table*}
 \caption{Setup and properties of the VLA observations (continued).}             
 \label{table:setup_2}
 \centering
 \begin{tabular}{l c c c c c c c} 
 \hline              
 Target & Waveband & Project & Observation & Beam size & Beam  & RMS & Measured \\
  & (GHz) & & date &  (\arcsec) & angle ($\degree$) & (mJy/beam) & flux (mJy) \\
 \hline                       
   UZ Tau E & Q (43) & 21B-267 & 13-12-21 & 0.18$\times$0.16 & -69 & 0.014 & 1.50 $\pm$ 0.10 \\  
   UZ Tau E & Ka (33) & 21B-267 & 18-12-21 & 0.20$\times$0.18 & -6 & 0.016 & 0.93 $\pm$ 0.07 \\  
   UZ Tau E & K (22) & 21B-267 & 21-12-21 & 0.34$\times$0.32 & 70 & 0.012 & 0.35 $\pm$ 0.03 \\  
   UZ Tau E & Ku (15) & 20A-373 & 17-2-20 & 1.61$\times$1.34 & -43 & 0.005 & 0.24 $\pm$ 0.02 \\  
   V409 Tau & Ka (33) & 20A-373 & 17-10-20 & 0.19$\times$0.17 & 78 & 0.007 & 0.22 $\pm$ 0.02 \\
   V409 Tau & Ku (15) & 20A-373 & 2-3-20 & 1.44$\times$1.32 & 68 & 0.004 & 0.04 $\pm$ 0.01 \\
   V409 Tau & C (6) & 20A-373 & 20-2-20 & 3.69$\times$3.04 & -12 & 0.008 & <0.03 \\
   V836 Tau & Ka (33) & 20A-373 & 12-8-20 & 0.25$\times$0.18 & -85 & 0.009 & 0.35 $\pm$ 0.03 \\
   V836 Tau & Ku (15) & 20A-373 & 8-3-20 & 1.61$\times$1.39 & -73 & 0.005 & 0.13 $\pm$ 0.02 \\
   V836 Tau & C (6) & 20A-373 & 25-2-20 & 3.14$\times$2.86 & 4 & 0.082 & <0.10 \\
 \hline 
 \end{tabular}
 \end{table*}

\end{document}